\documentclass[prd,twocolumn,showpacs,superscriptaddress,footinbib]{revtex4-1}
\usepackage{amsfonts}
\usepackage{amsmath}
\usepackage{amssymb}
\usepackage{bm}
\usepackage{dcolumn}
\usepackage{graphicx}
\usepackage{graphics}
\usepackage{latexsym}
\usepackage{rotating}
\usepackage[colorlinks=true]{hyperref}
\usepackage[all]{hypcap} 
\usepackage{xspace} 
\usepackage[usenames]{xcolor}
\usepackage{mathrsfs}
\usepackage{multirow}
\usepackage{soul}
\usepackage{ulem}
\definecolor {darkgreen}{rgb}{0.2,0.7,0.2}

\newcommand\be{\begin{equation}}
\newcommand\ee{\end{equation}}
\newcommand\bw{\begin{widetext}}
\newcommand\ew{\end{widetext}}

\newcommand{\bea}{\begin{eqnarray}}
\newcommand{\eea}{\end{eqnarray}}

\usepackage{cellspace} %
\def\apj{Astrophys. J. \ }%
\def\apjl{Astrophys. J. Lett. \ }%
\def\mnras{MNRAS}%
\begin{document}
\title{On the dynamical stability of quasi-toroidal differentially
  rotating neutron stars}

\author{Pedro L. Espino}
\affiliation{Department of Physics, University of Arizona, Tucson, AZ 
85721, USA}
\author{Vasileios Paschalidis}
\affiliation{Department of Physics, University of Arizona, Tucson, AZ 
85721, USA}
\affiliation{Department of Astronomy, University of Arizona, Tucson, 
  AZ 85721,USA}
\author{Thomas W. Baumgarte}
\affiliation{Department of Physics and Astronomy, Bowdoin College, 
Brunswick, ME 04011, USA}
\author{Stuart L. Shapiro}
\affiliation{Department of Physics, University of Illinois at Urbana-
Champaign, Urbana, Illinois 61801, USA}
\affiliation{Department of Astronomy and NCSA, University of Illinois 
at Urbana-Champaign, Urbana, Illinois 61801, USA}

\begin{abstract}
We investigate the dynamical stability of relativistic, differentially
rotating, quasi-toroidal models of neutron stars through
hydrodynamical simulations in full general relativity. We find that
all quasi-toroidal configurations studied in this work are dynamically
unstable against the growth of non-axisymmetric modes. Both one-arm
and bar mode instabilities grow during their evolution. We find that
very high rest mass configurations collapse to form black holes. Our
calculations suggest that configurations whose rest mass is less than
the binary neutron star threshold mass for prompt collapse to black
hole transition dynamically to spheroidal, differentially rotating
stars that are dynamically stable, but secularly unstable. Our study
shows that the existence of extreme quasi-toroidal neutron star
equilibrium solutions does not imply that long-lived binary neutron
star merger remnants can be much more massive than previously
found. Finally, we find models that are initially supra-Kerr
($J/M^2>1$) and undergo catastrophic collapse on a dynamical
timescale, in contrast to what was found in earlier works. However,
cosmic censorship is respected in all of our cases. Our work
explicitly demonstrates that exceeding the Kerr bound in rotating
neutron star models does not imply dynamical stability.

\end{abstract}

\date{\today}
\maketitle


\section{Introduction}
 Following the
first ever multi-messenger detection with gravitational waves (GWs) of
a binary neutron star (BNS) \cite{multimessGW170817}, there have been
a number of studies considering the stability of the merger remnant to
place constraints on the nuclear equation of state
\cite{Shibata:2017xdx, Ruiz:2017due, Margalit2017,
  Bauswein:2017vtn,Radice:2018xqa, Rezzolla:2017aly}. The exact nature
of the merger remnant is unknown, but it is possible that the remnant
was a hypermassive neutron star (HMNS)
\cite{Paschalidis:2016agf,Gill:2019bvq, Baiotti:2016qnr,
  TheLIGOScientific:2017qsa}.  HMNSs are differentially rotating stars
with rest mass greater than that allowed by uniform rotation
\cite{HypermassiveNSorig} (i.e, the supramassive limit \cite{CST92}).
The solution space of differentially rotating neutron stars has been
studied in great detail for polytropes of varying polytropic index
\cite{Ansorg:2009, Gondek-Rosinska:2016tzy, Studzinska:2016ofb} (see
also~\cite{Paschalidis:2016vmz} for a review), and recently for
realistic nuclear equations of state (EOSs) \cite{Espino:2019ebx},
strange quark star EOSs \cite{Szkudlared2019, Zhou:2019hyy}, and
hybrid hadron-quark EOSs \cite{Bozzola:2019tit}. The solution space of
differentially rotating neutron stars in equilibrium includes
configurations that can support more than twice the maximum
supportable rest mass by a non-rotating model with the same EOS, i.e,
the Tolman-Oppenheimer-Volkoff (TOV) limit. There are even models that
can support more than twice the supramassive limit mass with the same
EOS. Such stars are highly unlikely to form following BNS
mergers~\cite{Espino:2019ebx}. These extreme, differentially rotating
configurations tend to be quasi-toroidal, i.e., equilibria where the
maximum energy (or rest mass) density of the fluid does not occur at
the center of mass of the configuration, but in a ring around
it. Quasi-toroidal configurations have so far been found only when
differential rotation is allowed.

Differentially rotating massive neutron stars naturally arise as
remnants of BNS mergers.  Recent numerical simulations have shown that
quasi-toroidal HMNSs can form following a BNS merger \cite{PEFS2016,
  Zhang:2017fsy}. There are also simulations that find a double core
structure (see, e.g., \cite{Paschalidis:2012ff, Stergioulas:2011gd}
and \cite{Baiotti:2016qnr} for a review on different types of BNS
merger remnants). Despite this possibility, most relativistic
simulations of isolated stars modeling BNS merger remnants have
focused on spheroidal configurations (see \cite{Paschalidis:2016vmz}
for a review). Dynamical simulations of HMNSs suggest that specific
features can arise during the evolution, such as the one-arm
instability \cite{Centrella2001, Saijo2003, Watts2005,Ou2006,
  Saijo2006, Radice:2016gym, Lehner:2016wjg}, other non-axisymmetric
instabilities~\cite{Corvino2010}, the bar mode instability
\cite{Shibata:2000jt, HypermassiveNSorig}, and the low-$T/|W|$
instability \cite{Shibata:2002mr, Shibata:2003yj, Watts2005,
  Saijo2006, CerdaDuran:2007yw, Passamonti:2014eea, Saijo:2016vcd,
  Yoshida:2016kol}.  Recent work has also considered the dynamical
stability of differentially rotating, spheroidal stars based on
approximate turning points \cite{Bozzola:2017qbu, Weih:2017mcw}.
Knowing whether such HMNSs are dynamically stable or unstable can
inform us about the most massive remnants that may form following a
BNS merger, how long such remnants may live for, the properties of the
black hole (BH) that forms when they collapse, and the subsequent
electromagnetic signatures that accompany the GWs. Isolated stars can
also help probe theoretical aspects of gravitation such as cosmic
censorship and the ergoregion instability (see,
e.g.,~\cite{1978CMaPh..63..243F,1978RSPSA.364..211C,1996MNRAS.282..580Y,Cardoso:2007az}).

In this paper we examine the dynamical stability of quasi-toroidal,
differentially rotating neutron stars modeled as $\Gamma=2$
polytropes. In \cite{Gondek-Rosinska:2016tzy,Studzinska:2016ofb}
differentially rotating models that can support up to four times the
TOV limit rest mass were found. In fact, there exists a continuum of
quasi-toroidal configurations that can support a range of masses.  The
reference mass in our study is the BNS threshold mass for prompt
collapse to BH, which is 1.3-1.7 times the TOV limit mass
\cite{Shibata:2003ga, ST, Bauswein:2013jpa, Bauswein:2014qla,
  Koppel:2019pys} depending on the equation of state (note that the
threshold mass refers to the pre-merger binary total rest mass).  We
consider equilibrium configurations that are above and below this
threshold. Based on the fact that such high-mass equilibrium
configurations exist, one might assume that long-lived BNS merger
remnants much more massive than the BNS threshold mass for prompt
collapse could possibly arise. Most of these extremely massive
differentially rotating configurations are highly quasi-toroidal and
their dynamical stability has never been tested before. If extreme
quasi-toroidal configurations are viable long-lived BNS merger
remnants, they should be stable on a dynamical timescale.

Here we initiate a study of the dynamical stability of several
quasi-toroidal configurations that have rest masses ranging from
astrophysically relevant values ($\sim 1.4$ times the TOV limit rest
mass) to extreme, likely astrophysically irrelevant values ($\sim 4.0$
times the TOV limit rest mass). We perform hydrodynamic simulations in
full 3+1 general relativity of these configurations, and find that all
quasi-toroidal configurations we investigate are dynamically unstable
against the development of non-axisymmetric modes. Both a one-arm and
a bar mode grow during the evolution. We find that BH formation on a
dynamical timescale is the outcome of configurations (quasi-toroidal
or spheroidal) with rest mass exceeding the BNS threshold mass for
prompt collapse to BHs. On the other hand, one of our quasi-toroidal
configurations with rest mass less than the BNS threshold mass for
prompt collapse transitions dynamically to a differentially rotating,
{\it spheroidal} configuration that is dynamically stable, but
secularly unstable.

Our work shows that the existence of the massive, extreme
quasi-toroidal neutron star solutions that were recently found in the
literature does not imply that dynamically stable BNS merger remnants
can exist with masses much larger than the BNS threshold mass for
prompt collapse to a BH. Finally, several of the models we study are
initially supra-Kerr ($J/M^2>1$), yet they undergo catastrophic
collapse on a dynamical timescale. This result is in contrast to what
was found in earlier work \cite{Giacomazzo:2011cv}, where dynamically
unstable, differentially rotating supra-Kerr models of $\Gamma=2$
neutron stars could not be found, and supra-Kerr models could only be
induced to collapse through severe pressure depletion. However, cosmic
censorship is respected in all of our cases. Our study explicitly
shows that exceeding the Kerr bound initially does not imply dynamical
stability of a rotating neutron star configuration.

The rest of the paper is structured as follows. In 
Sec.~\ref{sec:basic_equations}, we review the properties of the solution
space of differentially rotating stars, and detail the properties of
the equilibrium configurations we adopt as initial data for our
simulations. In Sec.~\ref{sec:methods}, we briefly describe the set
of initial perturbations we consider, our evolution code, and the
diagnostics we employ in our analysis.  In Sec.~\ref{sec:res} we
describe the results of our simulations. In 
Sec.~\ref{sec:discussion} we discuss our findings in connection with key
global properties of the initial configurations and the final state of
our quasi-toroidal configuration that does not collapse to a BH. In
Sec.~\ref{sec:conclusion} we present out conclusions and discuss
possible future avenues of investigation. Throughout this paper we
adopt geometrized units, where $c = G = 1$ (where $c$ is the speed of
light in vacuum and $G$ the gravitational constant). We commonly
designate the TOV limit rest mass as $M_{0,\rm max}^{\rm TOV}$, and
the gravitational [or Arnowitt-Deser-Misner mass (ADM)] mass as $M$.

\section{Solution space of Differentially Rotating Stars and Initial Equilibria}
\label{sec:basic_equations}

The spacetime of stationary, axisymmetric, rotating neutron star
equilibria is described in spherical polar coordinates $r$ and $\theta$ by
the line element \cite{CST94a} (see also \cite{Paschalidis:2016vmz}
for a review of other line elements used in the literature),
\begin{equation}
  \label{eq:ds}
ds^2 = -e^{\gamma+\rho}dt^2+e^{2\alpha}(dr^2+r^2d\theta^2)+e^{\gamma-\rho}r^2\sin^2\theta(d\phi-\omega dt)^2,
\end{equation}
where $\gamma(r,\theta), \rho(r,\theta), \alpha(r,\theta)$ and
$\omega(r,\theta)$ are the metric potentials determined by the
solution of the Einstein equations coupled to the equation of
hydrostationary equilibrium for perfect fluids. The matter is modeled
as a perfect fluid whose stress-energy tensor is given by
\begin{equation}
T^{ab} = \rho_0 h u^a u^b + p g^{ab},
\end{equation}
where $u^a$, $\rho_0$ and $p$ are the fluid four velocity, rest mass
density, and pressure, respectively; $h$ is the specific enthalpy,
given by
\begin{equation}
h = 1 + \epsilon + \dfrac{p}{\rho_0},
\end{equation}
with $\epsilon$ the specific internal energy.
To close the system of equations an EOS must be supplied. In this work
we focus on polytropic EOSs which are described by
\begin{equation}\label{eq:poly}
p = \kappa \rho_0 ^\Gamma,
\end{equation}
where $\kappa$ is the polytropic constant and $\Gamma$ is the
adiabatic index. In particular, we consider stars that are described
by $\Gamma=2$. We also adopt polytropic units (equivalent to setting
$\kappa=1$) unless otherwise noted.

The integrability condition on the equation of hydrostationary
equilibrium enforces that the specific angular momentum ($j=
u^tu_\phi$) be a function of the angular velocity, i.e., $j =
u^tu_\phi = F(\Omega)$.  By choosing $F(\Omega)$, we specify a rotation
law for the matter. Here we work with the Komatsu-Eriguchi-Hachisu
(KEH) rotation law \cite{Komatsu:1989zz},
\begin{equation}
\label{eq:komatsu_rotlaw}
F(\Omega) = A^2 (\Omega_c - \Omega),
\end{equation}
where $\Omega_c$ is the angular velocity at the pole. The parameter
$A$ in Eq.~\eqref{eq:komatsu_rotlaw} has units of length, and
parametrizes the length scale over which the angular velocity changes
in the star. As in previous studies, we work with a rescaled version of $A$
given by
\begin{equation}
\label{eq:Am1}
\hat{A}^{-1} = \dfrac{r_e}{A},
\end{equation}
where $r_e$ is the equatorial radius of the configuration. The
 parameter $\hat{A}^{-1}$ (which we refer to as the degree of
differential rotation) lies in the range $0 \leq \hat{A}^{-1} < \infty$, with $\hat{A}^{-1} =
0$ corresponding to uniform rotation.

\begin{figure}[t]
\begin{tabular}{c }
\includegraphics[width=8.5cm]{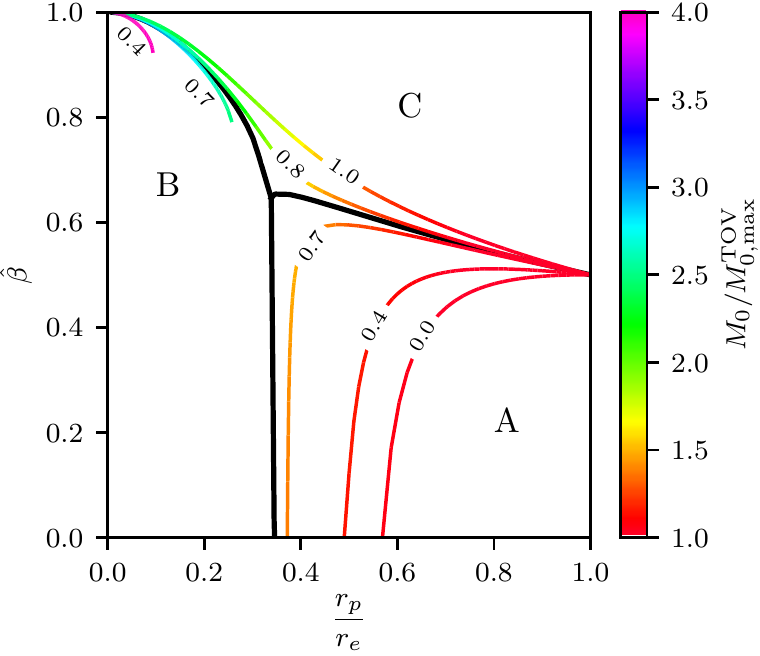} 
\end{tabular}
\caption{Projection of the solution space for a $\Gamma=2$ polytrope
  in the $(r_p/r_e, \hat{\beta})$ plane for a fixed value of the 
  stellar maximum energy density
  $\epsilon_{\rm max}=0.12$ (in polytropic units) and several values
  of the degree of differential rotation $\hat{A}^{-1}$.  Each
  sequence is labeled by $\hat{A}^{-1}$. The bold black line 
  corresponds to the
  critical value of the degree of differential rotation
  $\hat{A}^{-1}_{\rm crit} = 0.75904$ \cite{Ansorg:2009}, which here
  divides the solution space into three regions corresponding to type
  A (bottom right), B (left), and C (top) models. The color bar
  corresponds to the rest mass of models along each sequence of fixed
  $\hat{A}^{-1},$ normalized by the maximum rest mass for a
  non-rotating model (i.e the TOV limit rest mass) $M_{0,\rm max}^{\rm
    TOV}$. }
\label{fig:Beta_rp}
\end{figure}

Under the assumption of the KEH rotation law, it was shown in
\cite{Ansorg:2009, Gondek-Rosinska:2016tzy, Studzinska:2016ofb} that
the solution space of differentially rotating, polytropic neutron
stars can generally be divided into four classes. It was recently also
shown that at least three of these classes also exist for realistic
EOSs \cite{Espino:2019ebx} and for hybrid EOSs
\cite{Bozzola:2019tit}. Each solution type may be characterized by 
specific ranges of the values in the quadruplet $\left( \epsilon_{\rm
  max}, r_p/r_e, \hat{A}^{-1}, \hat{\beta} \right)$, where
$\epsilon_{\rm max}$ is the maximum energy density in the star,
$r_p/r_e$ is the ratio of polar to equatorial radius (which
parametrizes the angular velocity at the center of the star), and
$\hat{\beta}$ is the mass-shedding parameter which measures how close
to the Keplerian limit the configuration is. The parameter
$\hat{\beta}$ is defined as \cite{Ansorg:2009}
\begin{equation}
\hat{\beta} = \dfrac{\beta}{1 + \beta},
\end{equation}
where
\begin{equation}
\beta = - \left(\dfrac{r_e}{r_p}\right)^2 \dfrac{\mathrm{d}(z_b^2)}{\mathrm{d}(\varpi^2)} \Bigg\vert_{\varpi = r_e}, 
\end{equation}
with $\varpi = r \sin(\theta)$ and $z=r\cos(\theta)$ the usual
cylindrical coordinates, and $z=z_b$ describing the surface of the star. The mass-shedding parameter takes on several
limiting values depending on the shape of the configuration. 
Configurations at the mass-shedding limit have
$\mathrm{d}(z_b^2)/\mathrm{d}(\varpi^2)\vert_{\varpi = r_e}=0$ and 
hence $\hat{\beta} = 0$; spherical (non-rotating) models correspond 
to $r_p/r_e=1$, $\mathrm{d}(z_b^2)/\mathrm{d}(\varpi^2)\vert_{\varpi =
  r_e}=-1$, and hence $\hat{\beta} = \frac{1}{2}$; quasi-toroids
correspond to $r_p/r_e\rightarrow 0$, and $\hat{\beta}\rightarrow 1$.
In this work we focus on configurations with small but non-zero values
of $r_p/r_e$, which are the quasi-toroids.

\begin{table*}[htb]
  \centering
  \caption{Properties of the equilibrium models considered in this
    work. For each model we list the model label/model type, the
    dimensionless spin parameter $J/M^2$, the central period divided
    by the ADM mass $T_c/M$, the rest mass $M_0$ in units of $M_{0,\rm
      max} ^{\rm TOV}$ (the TOV limit rest mass), the ADM mass $M$ in
    units of $M_{\rm max}^{\rm TOV}$ (the TOV limit gravitational
    mass), the compactness $C=M/R_c$ (with $R_c$ the circumferential
    radius at the equator), the ratio of kinetic to gravitational
    potential energy $T/|W|$, the maximum energy density
    $\epsilon_{\rm max}$ in units of $\epsilon_{\rm max}^{\rm TOV}$
    (the maximum energy density of the TOV limit configuration), the
    ratio of polar to equatorial radius $r_p/r_e$, the degree of
    differential rotation $\hat{A}^{-1}$, and the mass-shedding
    parameter $\hat{\beta}$.\label{tab:maxmass_prop}}
  \begin{tabular}{l c c c c c c c c c c c c c}\hline  \hline
Model & $J/M^2$ & $T_c/M $ & $M_0/M_{0,\rm max}^{\rm TOV}$ & $M/M^{\rm TOV}_{\rm max}$ & $C$ & $T/|W|$ & $\epsilon_{\rm max}/\epsilon_{\rm max}^{\rm TOV}$ & $r_p/r_e$ & $\hat{A}^{-1}$ & $\hat{\beta}$\\ \hline
A & 0.89 & 27.95 & 1.63 & 1.66 & 0.22 & 0.22 & 0.74 & 0.35 & 0.7 & 0.66 \\
B & 1.07 & 24.78 & 3.79 & 3.73 & 0.28 & 0.33 & 0.21 & 0.035 & 0.4 & 0.99\\
C & 1.02 & 21.97 & 2.57 & 2.59 & 0.25 & 0.29 & 0.23 & 0.005 & 0.8 & 0.99\\
$\rm B_{low}$ & 1.56 & 144.67 & 1.36 & 1.47 & 0.09 & 0.3 & 0.05 & 0.005 & 0.8 & 0.99 \\
$\rm C_{low}$ & 0.89 & 15.38 & 1.81 & 1.85 & 0.23 & 0.24 & 0.29 & 0.01 & 1.5 & 0.99\\ \hline
\end{tabular}
 \end{table*}

\begin{figure*}[htb]
\begin{tabular}{c }
\includegraphics[width=5.5cm]{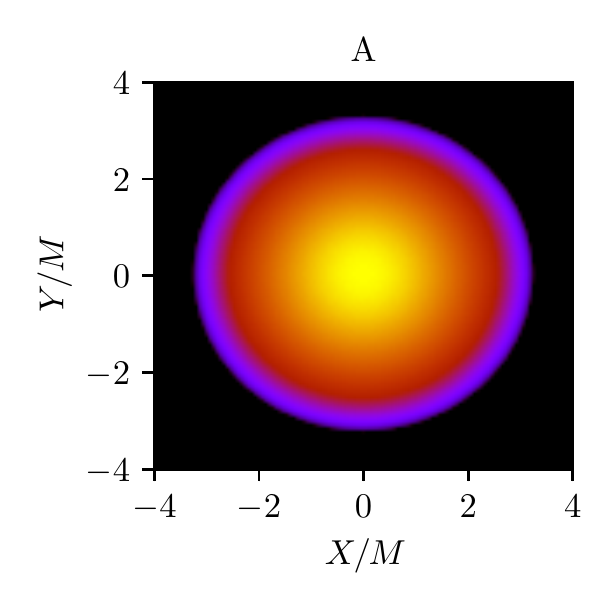}
\includegraphics[width=5.05cm]{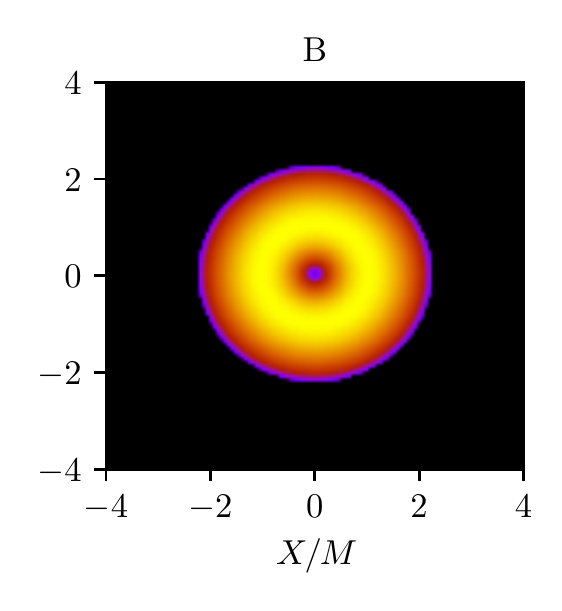}
\includegraphics[width=6.65cm]{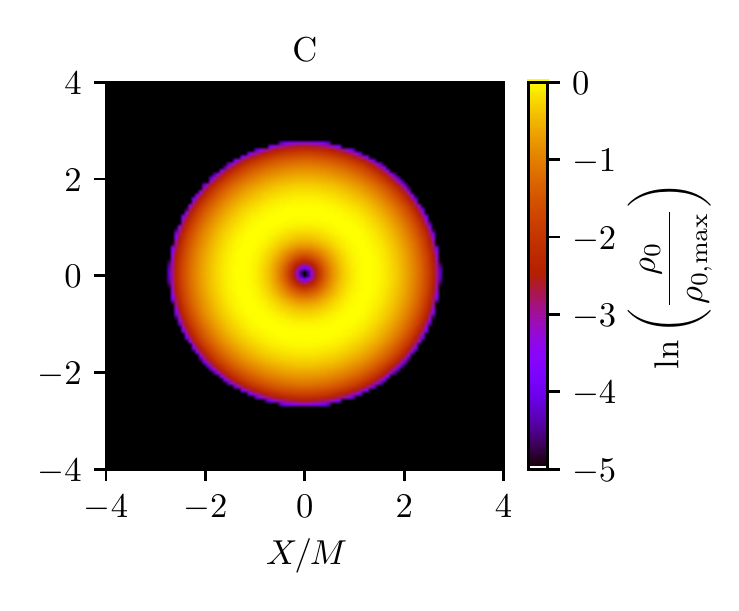}\\
\end{tabular}
\caption{Equatorial contours of the rest mass density $\rho_0$ at $t=0$
  for the maximum rest mass A, B, and C type models (left, middle, and
  right, respectively) in Table~\ref{tab:maxmass_prop}.  The color bar
  shows the value of the rest mass density scaled to the maximum value  on a logarithmic scale.
}
\label{fig:merid_t0}
\end{figure*}

In Fig.~\ref{fig:Beta_rp} we show a projection of the solution space
for a $\Gamma=2$ polytrope in the $(\hat{\beta},r_p/r_e)$ plane at a
fixed value of $\epsilon_{\rm max}$. The methods we adopt for building
these stars are the same as those developed in~\cite{Espino:2019ebx}.
The different lines in Fig.~\ref{fig:Beta_rp} correspond to sequences
of constant $\hat{A}^{-1}$ [or contours of the function
  $\hat{A}^{-1}(\hat{\beta},r_p/r_e)$].  Each curve is labeled by its
corresponding value of $\hat{A}^{-1}$.  The solid black line
corresponds to the critical degree of differential rotation
$\hat{A}^{-1}_{\rm crit}$ which divides the solution space in four
regions, each corresponding to one of the four solution types. In this
study we will focus on the solution types that we were able to build
with the Cook et al. code \cite{CST92, CST94a, CST94b}, namely the type A, B,
and C solutions detailed in \cite{Ansorg:2009}. Since we were unable
to construct type D models, the $\hat{A}^{-1}_{\rm crit}$ curve
divides the plot in Fig.~\ref{fig:Beta_rp} into only three regions (see 
Fig. 2 in \cite{Ansorg:2009} for an example of the complete 
solution space for a $\Gamma=2$ polytrope with 
$\epsilon_{\rm max} =0.12$).  The type A solutions consist of 
spheroidal models and correspond to relatively low degrees
of differential rotation $\hat{A}^{-1} < \hat{A}^{-1}_{\rm
  crit}$. Spinning these stars up, i.e., decreasing $r_p/r_e$, results
in mass-shedding. Type A solutions reside in the lower right part of
Fig.~\ref{fig:Beta_rp}, where one end of the sequences is located at
$\hat{\beta} = 0.5$ (corresponding to a spherical model) and the other
is located at $\hat{\beta} = 0$ (corresponding to mass-shedding). Type
B solutions are related to type A models, in that they may exist for
the same values of $\hat{A}^{-1} < \hat{A}^{-1}_{\rm crit}$, but they
are quasi-toroidal. The type B solutions correspond to the left side
of Fig.~\ref{fig:Beta_rp}; these sequences have one end located at
$\hat{\beta} = 1$ (corresponding to a toroidal model) and the other
located at $\hat{\beta} = 0$ (corresponding to mass-shedding).  We
were unable to construct type B stars pinched at the equator, which
results in type B sequences shown in Fig.~\ref{fig:Beta_rp} that do
not terminate at the mass-shedding limit. Type B solutions are the
most massive among the four types, as also depicted by the color bar
in Fig.~\ref{fig:Beta_rp}. The type C sequences include both
spheroidal and quasi-toroidal models. Nevertheless, the most massive
models in this class of stars tend to be extremely close to a toroidal
topology. The fourth solution class, type D, is also quasi-toroidal
(see~\cite{Ansorg:2009} for more details).  For each of the solution
types considered, we built the maximum rest mass models found in
\cite{Ansorg:2009}.

We searched the solution space for the maximum rest mass models by
building sequences of solutions at constant $\epsilon_{\rm max}$ as
detailed in \cite{Espino:2019ebx}.  We present relevant properties of
these maximum rest mass models in Table~\ref{tab:maxmass_prop}.  For
the two quasi-toroidal solution types, we also consider lower mass
equilibria (labeled $\rm B_{low}$ and $\rm C_{low}$) to probe the role
that the rest mass plays on stability. As shown in 
Table~\ref{tab:maxmass_prop} the maximum mass type B models are the most
massive ones supporting up to four times the TOV limit rest
mass. Equatorial contours of the rest mass density of the most massive
A, B, and C type models in Table~\ref{tab:maxmass_prop} are shown in
Fig.~\ref{fig:merid_t0}. Figure~\ref{fig:merid_t0} shows the
quasi-toroidal nature of the B and C models, where an under-dense
region exists at the geometric center of the configuration, and the
maximum rest mass density is located in a ring around the center.
These models, along with the $\rm B_{low}$ and $\rm C_{low}$ models listed in
Table~\ref{tab:maxmass_prop}, represent our initial equilibria.

\section{Methods}
\label{sec:methods}

We evolve the initial data presented in 
Sec.~\ref{sec:basic_equations} using the well-tested Illinois GRMHD code
\cite{Duez:2005sg, Etienne:2015cea} which operates within the {\tt
  Cactus} infrastructure \cite{Cactus} and uses {\tt
  Carpet}~\cite{Schnetter:2003rb, Schnetter:2006pg} for mesh
refinement. Illinois GRMHD solves the Einstein equations within the
ADM 3+1 framework and evolves the spacetime using the BSSN
formulation of the Einstein equations \cite{Shibata:1995we,
  Baumgarte:1998te}. Our gauge choice employs 1+log time slicing for
the lapse $\alpha$ \cite{Bona:1994dr}, and the ``Gamma-freezing''
condition for the shift $\beta^i$ cast in first order form
\cite{Alcubierre:2001vm, Alcubierre:2002kk} (see also Eqs. (2)-(4) in
\cite{Etienne:2007jg}).  We use the \texttt{MoL} thorn to solve the
equations in time by use of a fourth order Runge-Kutta scheme with the
Courant factor set to 0.5.  We ignore magnetic fields, and the
equations of hydrodynamics are solved in conservation-law form
adopting the high-resolution shock-capturing methods described
in~\cite{Etienne:2011re,Etienne:2010ui}. To close the evolution
system, an EOS needs to be provided. We adopt a $\Gamma$-law EOS $P =
(\Gamma-1)\rho_0 \epsilon$, with $\Gamma=2$, for the evolution.
  
\subsection{Grid hierarchy}  
\label{subsec:grid}
Our fixed mesh refinement grid hierarchy consists of nested cubes with
7 refinement levels. The finest level half-side length is set to
$r_1\approx 1.25R_{\rm NS}$, where $R_{\rm NS}$ is the neutron star
coordinate equatorial radius. Thus, the entire star is covered by the
finest level.  The half-side length of refinement level $n$ is set to
$r_n = 2^{(n-1)} r_1$ (where $n=1$ corresponds to the finest level and
n=7 to the coarsest one).  We set the spatial resolution on the finest
level to $dx_1 = M/20$ in order to capture BH properties should a BH
form following collapse of the star, where $M$ is the ADM mass of the
initial configuration. Each subsequent refinement level has half the
resolution of the previous. Therefore, the resolution of refinement
level $n$ is given by $dx_n = 2^{(n-1)}dx_1$. Cartesian coordinates
are adopted, and equal resolution is chosen for the $x$, $y$, and $z$
directions. We impose reflection symmetry across the equatorial plane,
such that our grid extent in the $z$ direction is $0\leq z \leq
80R_{\rm NS}$. We do not impose a $\pi-$rotational symmetry, so that
odd-number non-axisymmetric modes are not artificially suppressed
\cite{Paschalidis:2015mla}.  In the type A and type B cases we also
performed simulations at 1.2 times and 1.5 times the canonical
resolution. The Cook et al. code uses spherical coordinates, whereas
Illinois GRMHD uses Cartesian coordinates. To avoid coordinate
singularities in transforming the initial data from spherical to
Cartesian coordinates we shift our Cartesian coordinates in the $y$
direction by a small amount to avoid the origin of the coordinate
system. In Appendix~\ref{app:high_res} we investigate the effects of
grid resolution and the y-coordinate shift on our results.

\subsection{Initial perturbations}
\label{sec:init_pert}
Each of the initial configurations presented in 
Table~\ref{tab:maxmass_prop} is evolved without and with initial
perturbations. We consider three types of perturbations: a) we evolve
the initial data after exciting a quasi-radial perturbation in the
star. We achieve this by locally depleting the pressure by $0.5\%$
everywhere in the star at $t=0$; we also excite b) one-arm ($m=1$),
and c) bar-mode ($m=2$) non-axisymmetric rest mass density
perturbations of the form \cite{Zink:2006qa}
\begin{equation}\label{eq:dens_pert}
\rho_0 \longrightarrow \rho_0 \left( 1 + \dfrac{B \varpi \sin(m\phi)}{r_e} \right),
\end{equation}
where $r_e$ is the stellar coordinate equatorial radius, $\phi$ is the
azimuthal coordinate, and $B$ is the perturbation amplitude. We excite
only one perturbation per evolution to determine the role that each
mode plays. Given that we have 5 configurations in
Table~\ref{tab:maxmass_prop}, and 4 types of evolutions, we have a
total of 20 cases in our study. In all of the evolutions considered,
we set $B=0.5\%$. Note that $0.5\%$ is the maximum perturbation near
the edge of the star. Near the location of the maximum density this is
reduced to $\sim 0.1\%$.  We have checked that the amplitude of our
initial perturbations is small enough that truncation error dominates
the initial constraint violations. Therefore, we do not resolve the
constraints after applying the perturbation.

\subsection{Diagnostics}
\label{subsec:diagnostics}
We use several diagnostics during the evolution to test for stability
against collapse, assess non-axisymmetric mode growth, measure black
hole properties, and extract gravitational waves (GWs).  We calculate
the maximum of the rest mass density as a function of time
$\rho_{0,\rm max}(t)$ to determine whether the configuration is
undergoing collapse.  The ``collapse'' of the lapse function is also used
as an indicator for BH formation.  We locate BH apparent
horizons (AH) with the \texttt{AHFinderDirect} thorn
\cite{Thornburg:2003sf}. The \texttt{AHFinderDirect} thorn provides
the BH irreducible mass as well as the equatorial and meridional AH
circumferences. The ratio $C_r$ of the meridional circumference to
that of the equator can be used to provide a good approximation to the
BH dimensionless spin, for which we employ the approximating formula
of \cite{Alcubierre2004}
\begin{equation}\label{eq:spin_circ}
a_{\rm BH} = \sqrt{1 - (2.55C_r - 1.55)^2}.
\end{equation}
This formula is derived for a Kerr spacetime, and is applicable to the
final black hole as the spacetime approaches the Kerr solution at late
times.

We compute the volume-integrated azimuthal density mode decomposition,
given by~\cite{Paschalidis:2015mla,PEFS2016,East:2016zvv}
\begin{equation}\label{eq:densmode_decomp}
C_m = \int\sqrt{-g} \rho_0u^0 e^{im\phi} d^3x,
\end{equation}
to test for growth of non-axisymmetric modes.  Note that $C_0$ is the
total rest mass of the configuration. Note also that Eq.
\eqref{eq:densmode_decomp} yields zero for $|m|> 0$, if the density,
velocity and metric fields are axisymmetric.

Equation \eqref{eq:densmode_decomp} is useful for a qualitative
understanding of the matter evolution, but does not provide a
gauge-invariant measure of non-axisymmetric modes. Therefore, we also
extract gravitational radiation to determine the growth of
non-axisymmetric modes during the evolution.  For this, we compute the
Newman-Penrose scalar $\Psi_4$ and decompose it into $s=-2$
spin-weighted spherical harmonics, to determine growth of
axisymmetric ($m=0$) and non-axisymmetric ($m\neq 0$) GW modes during evolution in a
gauge-invariant way. We denote the coefficients of this decomposition
as $\Psi_4^{l,m}$, and focus on the $l=2$, $m=0$, $m=1$, and $m=2$
modes in this work. We compute $|\Psi_4^{2,m}|$ in the wave zone. We
find that generally $C_m$ and $\Psi_4^{2,m}$ are consistent as
indicators of non-axisymmetric mode growth, i.e., the same
non-axisymmetric modes that are excited in $C_m$ are also excited in
$\Psi_4^{2,m}$. This is not unexpected because the decomposition in
both diagnostics carries the $e^{im\phi}$ term. We show these
diagnostics in Sec.~\ref{sec:res}.

Finally, we also monitor the $\rm L_2$ norm of the Hamiltonian
$||\mathcal{H}||$ and momentum $||\mathcal{M}||$ constraints via
Eqs. (40) and (41) in \cite{Etienne:2007jg}, which along with $\Psi_4$
we use to demonstrate convergence in Appendix~\ref{app:high_res}.

\begin{figure*}
\includegraphics[width=5.9cm]{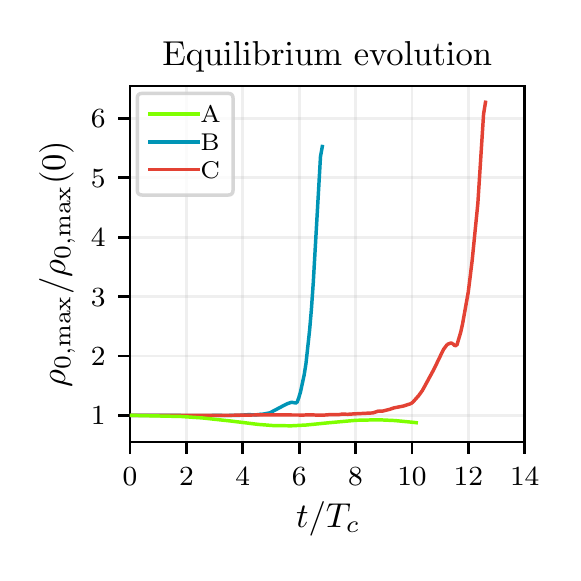}
\includegraphics[width=5.9cm]{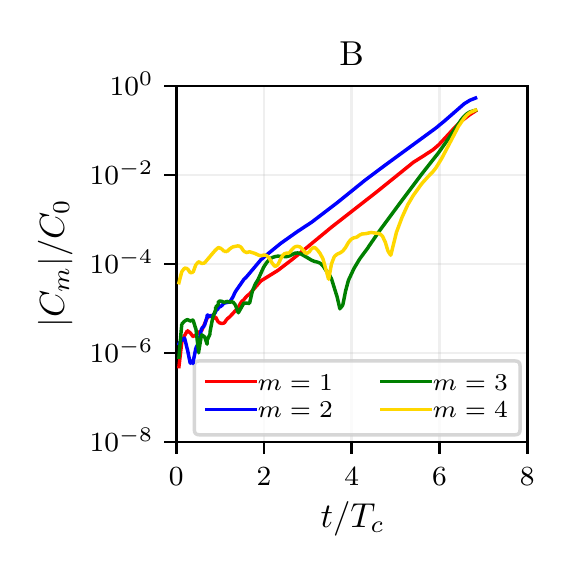}
\includegraphics[width=5.9cm]{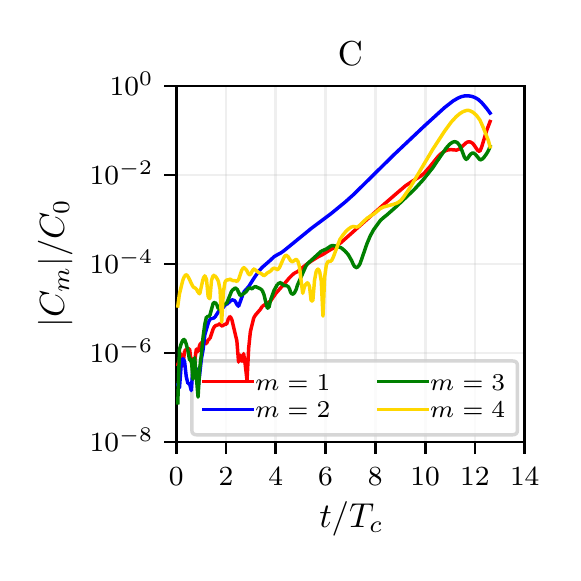}\\
\caption{Left panel: Maximum rest mass density as function of time for
  the A, B, and C models in the case of zero initial
  perturbations. The green, blue, and red lines correspond to the A,
  B, and C, models, respectively.  Center panel: evolution of the
  amplitude $|C_m|$ of non-axisymmetric density modes for the B model
  [$m=1\ (\rm{ red})$, $m=2\ (\rm{ blue})$, $m=3\ (\rm{ green})$,
    $m=4\ (\rm{ yellow})$] in the case of zero initial perturbations.
  Right panel: Same as the center panel but for the C model.}
\label{fig:Equi_densmode}
\end{figure*}

\begin{figure*}
\includegraphics[width=5.6cm]{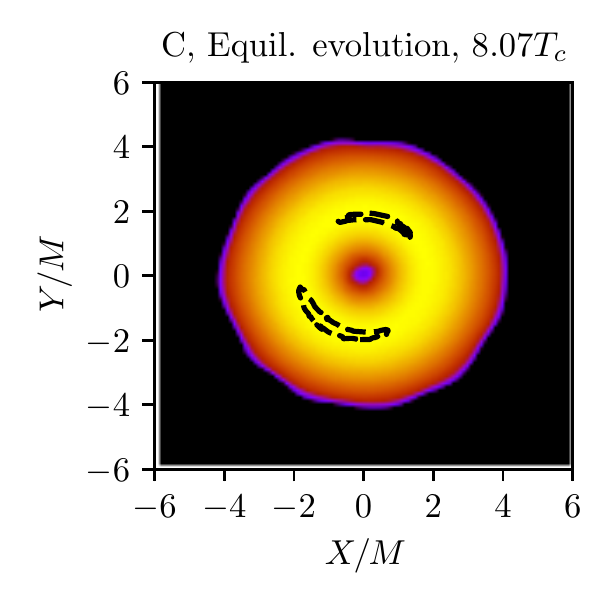}
\includegraphics[width=5.3cm]{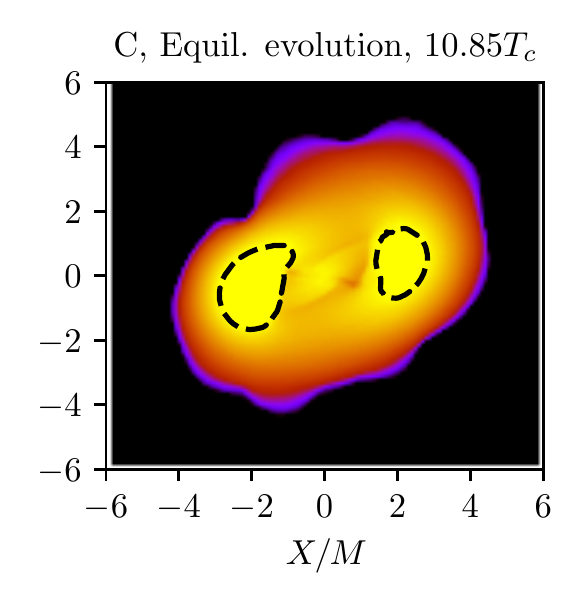}
\includegraphics[width=6.7cm]{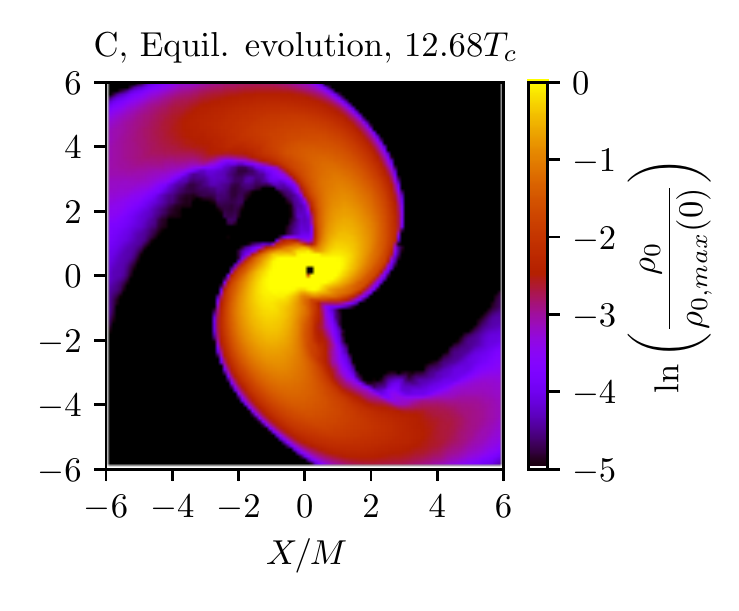}
\caption{Top panel: Snapshots of equatorial rest mass density contours $\rho_0$,
  scaled to the maximum value at the start of simulation 
  $ \rho_{0,\rm max}(0)$ for the C model under equilibrium evolution. 
 The dashed lines indicate the boundary of the regions within which 
 the rest mass density satisfies $\rho_0 \geq \rho_{0,\rm max}(0)$. 
 The left and center panels show the development of the model at 
 $t=8.07 T_c$ and $t=10.85 T_c$, respectively. 
 The right panel shows the state of 
 the model in the time near collapse at $t=12.68 T_c$.}
\label{fig:BC_equi_snapshots}
\end{figure*}

\section{Results}

\label{sec:res}
In this section we present the results from dynamical spacetime
simulations of the models listed in Table~\ref{tab:maxmass_prop} under
no initial perturbation and the three types of initial perturbations we described in
Sec.~\ref{sec:init_pert}.  For each model we scale the evolution time
by the initial period at the center of the configuration $T_c$, all
rest mass densities by the maximum at the start of simulation
$\rho_{0, \rm max}(0)$, and all density mode amplitudes by the
amplitude of the dominant $C_0$ mode. We generally find that the
quasi-toroidal models considered here are unstable to non-axisymmetric
mode growth on dynamical timescales. First, we discuss our study of
the most massive models, and subsequently we present the results from
the lower-mass models.

\subsection{Most massive A, B, C models}
\label{subsec:resABC}
Here we report our results for the most massive A, B, and C
models listed in Table~\ref{tab:maxmass_prop}, categorized by the type
of perturbation considered at the start of simulation.

\subsubsection{Evolution of equilibrium configurations without perturbation}
\label{subsec:Equil}

When evolving the initial equilibria without perturbation, we find
that all models except for model A are unstable to the growth of
non-axisymmetric modes on a dynamical timescale. We will often refer to
 evolutions without initial perturbation as ``equilibrium evolutions''.

In the left panel of Fig.~\ref{fig:Equi_densmode} we show the maximum
rest mass density $\rho_{0,\rm max}$ as a function of time.  We show
the evolution either up to collapse (for models which form BHs) or
until an approximately steady state has been reached (for
non-collapsing models). The figure shows that the equilibrium
evolution of the A model results in a $\sim 10\%$ oscillation of the
maximum rest mass density about the value $\rho_{0, \rm
  max}/\rho_{0,\rm max}(0) \approx 0.87$. In this case, we evolved the
A model until $t \approx 10 T_c$ and saw no sign of dynamically
unstable mode growth by the end of simulation. However, together with
the fact that this model is unstable to radial perturbations (as
discussed in the next sub-section), the oscillation of $\rho_{0, \rm
  max}$ by $\sim 10\%$ suggests that this model is only marginally
stable. By contrast, in models B and C the density grows slowly until
it reaches a point after which it increases rapidly and the
configurations undergo catastrophic collapse.

Although we do not excite any perturbations, seeded perturbations at
the level of truncation error grow such that all quasi-toroidal models
exhibit non-axisymmetric instabilities. This can be seen in the center
and right panels of Fig.~\ref{fig:Equi_densmode}, where we show the
azimuthal density mode decomposition, with mode amplitudes given by
Eq.~\eqref{eq:densmode_decomp}. We focus on the
$m=1,\ 2,\ 3,\ \text{and } 4$ modes. We find that the $m=1$ and $m=2$
modes grow at similar rates (exponentially with time) dominating over
the higher modes.  Nevertheless, the $m=2$ mode is dominant throughout
most of the evolution for the quasi-toroidal models. Model A does not
exhibit any growth of non-axisymmetric modes, and is not plotted here.
We find that the B and C configurations are especially unstable to
non-axisymmetric modes corresponding to the bar mode instability, even
though the $m=2$ mode was not explicitly excited at the start of
simulation.  The evolution of the cases developing strong $m=2$ modes
generally proceeds as in the dynamical bar mode instability
\cite{HypermassiveNSorig, Shibata:2000jt, Loffler2015}, but eventually
leads to catastrophic collapse.

In Fig.~\ref{fig:BC_equi_snapshots} we show density contours on the
equatorial plane that demonstrate how the dynamics of a strong $m=2$
mode proceeds in the C model under equilibrium evolution. First, two
over-dense regions develop in the ring of maximum density of the
quasi-toroids (left panel of Fig.~\ref{fig:BC_equi_snapshots}). Next,
as the two over-dense regions move apart, a typical high-density bar
develops, with the two over-dense regions forming the ``arms'' of the
bar (center panel of Fig.~\ref{fig:BC_equi_snapshots}). As the two
over-dense arms coalesce near the geometric center of the
configuration, the maximum rest mass density in the bar continues to
rise. For the most massive models, such as the B and C models, the
maximum density grows rapidly until complete gravitational
collapse ensues and a single BH forms near the center of mass (right
panel of Fig.~\ref{fig:BC_equi_snapshots}).

The value of $\rho_{0, \rm max} (t)$ also shows features which 
are consistent with a dominant $m=2$ mode in cases that 
develop $m=2$ non-axisymmetries in our study. As shown in the 
left panel of Fig.~\ref{fig:Equi_densmode}, $\rho_{0, \rm max}$ 
shows a local peak prior to collapse (the local peak is seen at 
$t\approx 5.75T_c$ for the B model and $t\approx 11T_c$ for the C 
model). This local maximum in time coincides with the saturation 
of the bar as it reaches maximum density.  After this brief 
saturation, the two over-dense arms of the bar mode bounce, 
launching shocks which lead to a momentary decrease in 
$\rho_{0,\rm max}$ that explains the ``dip'' in the evolution of 
the maximum rest mass density. Eventually, the over-dense arms 
coalesce near the geometric center, leading to a significant rise
in $\rho_{0, \rm max}$, and ultimately to catastrophic collapse.  
This ``double-peak'' feature is observed in many of our cases 
with a dominant $m=2$ mode.

\subsubsection{Pressure depletion}
\begin{figure*}
\includegraphics[width=5.9cm]{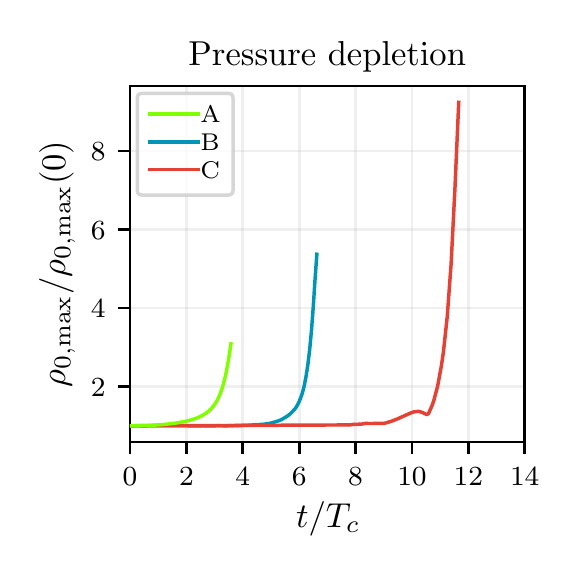} 
\includegraphics[width=5.9cm]{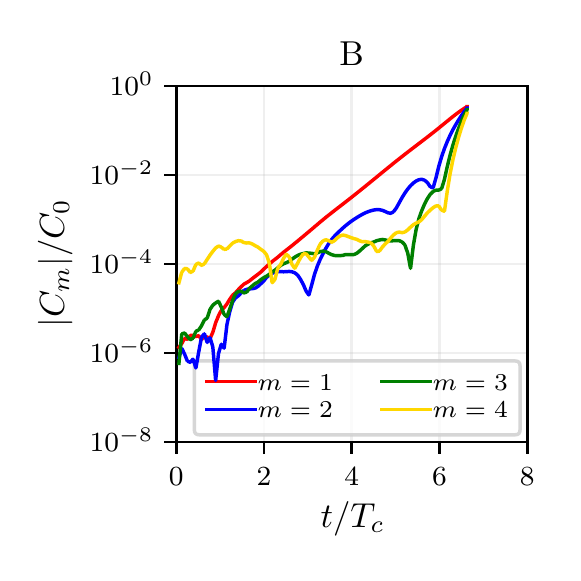} 
\includegraphics[width=5.9cm]{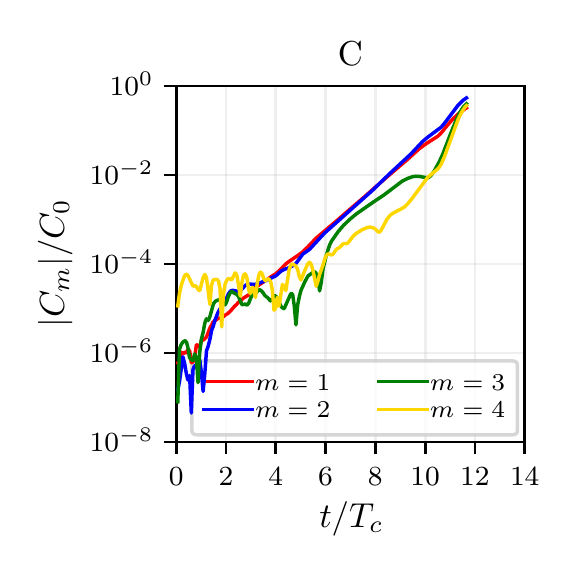}\\
\caption{
Same as Fig.~\ref{fig:Equi_densmode} but for the case of 
pressure depletion.}
\label{fig:Depl_densmode}
\end{figure*} 

In the left panel of Fig.~\ref{fig:Depl_densmode} we show the
evolution of the maximum rest mass density and the density mode
decomposition in the case of pressure depletion for the most massive
models.  The evolution of the maximum rest mass density is comparable
to the case of equilibrium evolution for the quasi-toroidal B and C
models, and they collapse practically on the same timescales as in the
evolution without perturbation. This suggests that these stars are not
quasi-radially unstable, but are unstable only to the development of
non-axisymmetric modes.

The quasi-radial ($m=0$) pressure depletion evolution was the only
type of evolution that resulted in collapse for the A model, which
indicates that on dynamical timescales it is unstable to collapse
against quasi-radial perturbations, but not against non-axisymmetric
ones.  Note that model A has rest mass of $M_0^{\rm A} = 1.8 M_{0,\rm
  max}^{\rm TOV}$, and the fact that it collapses to a BH following a
quasi-radial perturbation is coincidentally consistent with the
threshold mass of 1.65-1.75$M_{0,\rm max}^{\rm TOV}$ for prompt
collapse in the case of $\Gamma =2$ BNS mergers \cite{Shibata:2003ga}.

Compared to the equilibrium evolutions, in the case of pressure
depletion we observe a stronger $m=1$ density mode developing early in
the evolution of the most massive quasi-toroids, with the $m=1$
density mode comparable to the $m=2$ mode in the time leading up to
collapse or by the end of simulation.  The evolution in the case where
an $m=1$ density mode dominates in quasi-toroids generally proceeds as
follows: first, a single over-dense region develops somewhere in the
ring of maximum rest mass density in the quasi-toroid.  Next, this
single over-dense region develops a single arm, akin to the one-arm
instability \cite{Centrella2001,Saijo2003}.  For massive enough
configurations, the over-dense region continues to collapse until it
forms a BH.

The similarity in amplitude and growth times for the $m=1$ and $m=2$
density modes in the cases of equilibrium and pressure depletion
evolutions of the quasi-toroidal models suggests that whichever mode
is excited first, and with stronger amplitude, will dominate
throughout the evolution. We test this expectation in the following
sub-section.

\subsubsection{Non-axisymmetric perturbations}
\begin{figure*}
\includegraphics[width=5.9cm]{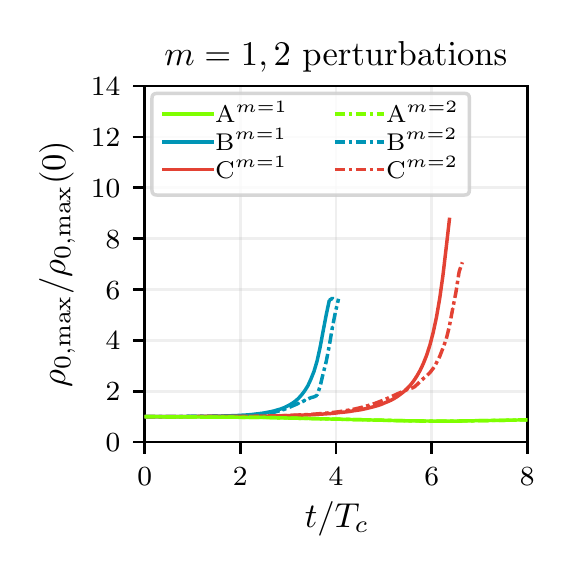} 
\includegraphics[width=5.9cm]{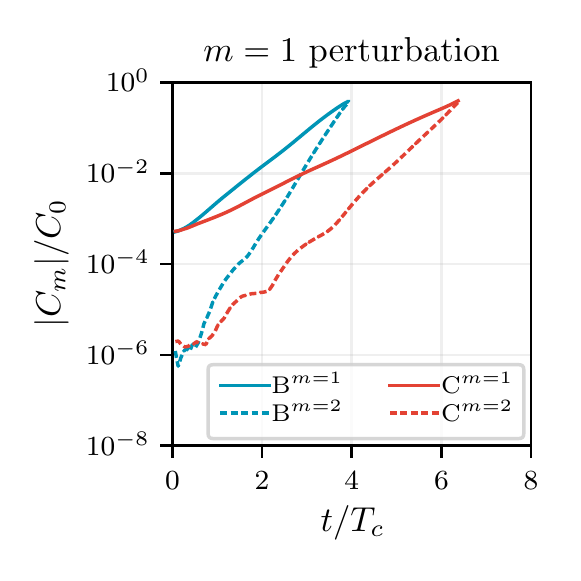} 
\includegraphics[width=5.9cm]{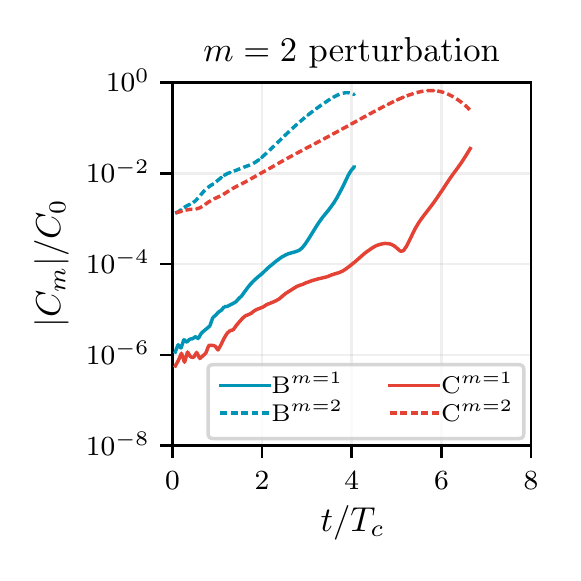} \\
\caption{Left panel: Maximum rest mass density as function of 
time for
  the A, B, and C models in the case of non-axisymmetric $m=1$ (solid
  lines) and $m=2$ (dash-dotted lines) initial rest mass density
  perturbations. The green, blue, and red lines correspond to the A,
  B, and C, models, respectively.  Center panel: evolution of the
  dominant $m=1$ (solid lines) and $m=2$ (dashed lines) non-axisymmetric density modes
  for the B (blue lines) and C (red lines) models in the case of
  an $m=1$ perturbation.  Right panel: Same as the center panel but
  for an $m=2$ perturbation.}
\label{fig:m1m2_densmode}
\end{figure*}

Under non-axisymmetric initial perturbations, the A model did not
collapse and evolved in a similar fashion to the equilibrium evolution
case [see the left panel of Fig.~\ref{fig:m1m2_densmode}, where we show the value of
$\rho_{0, \rm max}(t)$ for the A, B, and C models under both $m=1$
(solid lines) and $m=2$ (dash-dotted lines) initial perturbations]. Thus, we focus
the discussion on the quasi-toroidal models here.

\label{subsec:m1m2}
\begin{figure*}
\includegraphics[width=5.9cm]{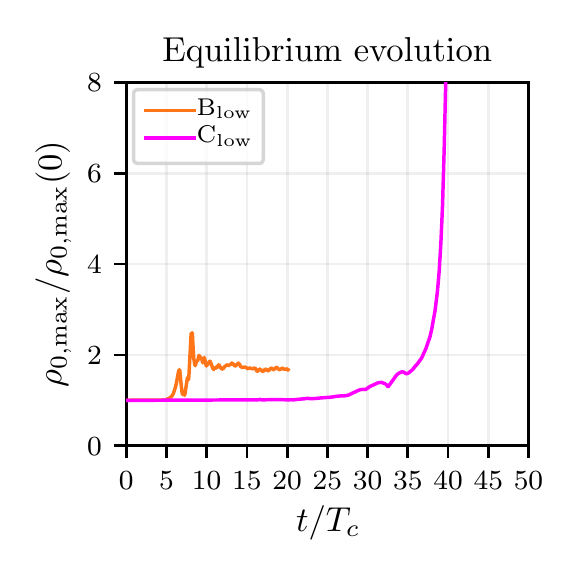} 
\includegraphics[width=5.9cm]{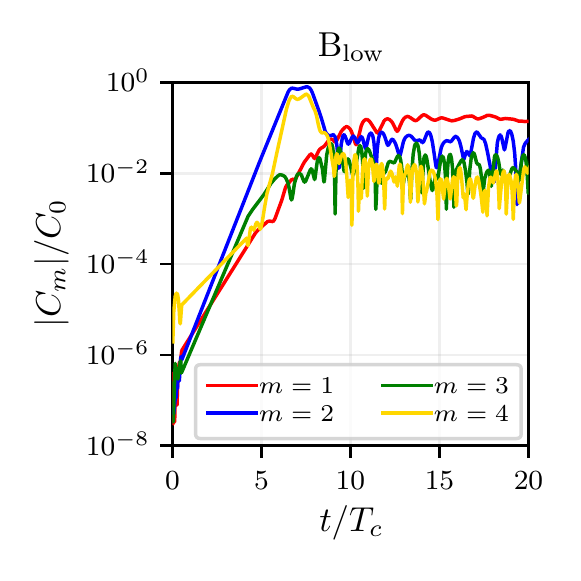} 
\includegraphics[width=5.9cm]{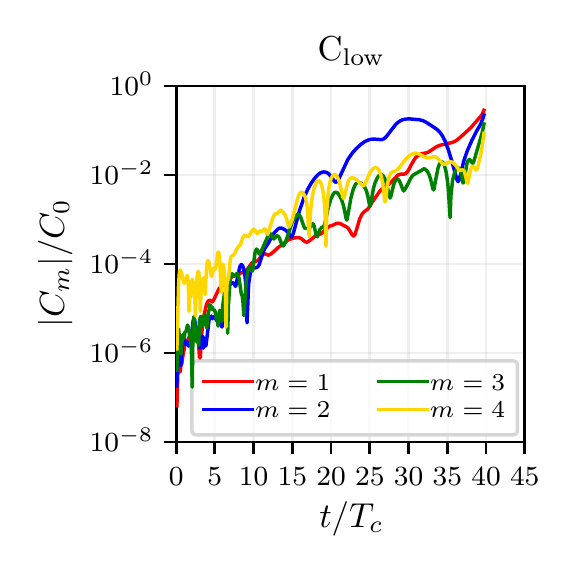}\\
\caption{Left panel: Maximum
  rest mass density as function of time for the $\rm B_{low}$ and $\rm
  C_{low}$ models in the case of zero initial perturbations. The
  orange and magenta lines correspond to the $\rm B_{low}$ and $\rm
  C_{low}$ models, respectively.  Center panel: evolution of
  non-axisymmetric density modes for the $\rm B_{low}$ model for the
  [$m=1\ (\rm{ red})$, $m=2\ (\rm{ blue})$, $m=3\ (\rm{ green})$,
    $m=4\ (\rm{ yellow})$] modes in the case of zero initial
  perturbations.  Right panel: Same as the center panel but for the
  $\rm C_{low}$ model. Note that model $\rm B_{low}$ has a central
  period which in units of $M$ is almost an order of magnitude longer
  than that of model $\rm C_{low}$. Thus, the evolution of model $\rm
  B_{low}$ is very long. It is the normalization with respect to $T_c$
  that makes it appear that this evolution is short.}
\label{fig:BClow_equi}
\end{figure*} 

In the cases discussed thus far, the dominant density modes during
evolution have been the $m=1$ and $m=2$ modes. To better understand
the features of evolution in the case of strong non-axisymmetric mode
growth, we excite initial perturbations of the form given in
Eq.~\eqref{eq:dens_pert} with $m=1$ or $m=2$. We find that the $m=1$
and $m=2$ density modes grow on very similar timescales for both the B
and C models, with the $m=1$ initial perturbation forcing slightly
earlier collapse than the $m=2$ initial perturbation (see the left panel of
Fig.~\ref{fig:m1m2_densmode}).

In the center and right panels of Fig.~\ref{fig:m1m2_densmode} we show
the density mode decomposition for the B (solid lines) and C (dashed
lines) models in the case of $m=1$ and $m=2$ initial perturbations,
respectively.  We focus on the evolution of two most dominant density modes
($m=1$ in red and $m=2$ in blue). Exciting an $m=1$ or $m=2$ mode at
the start of simulation ensures that the corresponding mode is
dominant throughout the evolution.  We observe that in the case where
an $m=1$ mode is initially explicitly excited, the amplitude of the
$m=2$ density mode becomes comparable to that of the $m=1$ mode near
collapse, but the $m=1$ density mode remains dominant. In the case
where an $m=2$ mode is initially excited, the $m=2$ density mode
remains significantly stronger than the $m=1$ mode even until collapse
(e.g, compare the center and right panels of
Fig.~\ref{fig:m1m2_densmode}). On the other hand, exciting an $m=1$ mode
initially tend to lead to faster collapse.

These results show that the one-arm instability is as important as the
bar mode instability in collapsing quasi-toroidal configurations, and
that the non-axisymmetric mode, which dominates early in the
evolution of an unstable quasi-toroidal configuration, is chiefly
responsible for its collapse.

\subsection{Low mass B and C models}
\label{subsec:Blow_res}

The most massive B and C models considered thus far have $M_{0, \rm
  max} > 2M_{0,\rm max}^{\rm TOV}$, making them unlikely models of BNS
merger remnants.  For this reason, we also consider the dynamical
stability of models with rest mass that could represent the total rest
mass of BNSs. In this section we present our results for the low mass
type B and C models (which we refer to as $\rm B_{low}$ and $\rm
C_{low}$) listed in Table~\ref{tab:maxmass_prop} to study the effect
of total mass on the stability of quasi-toroidal neutron stars.

\subsubsection{Evolution of equilibrium configurations without perturbation}
\begin{figure*}
\includegraphics[width=5.9cm]{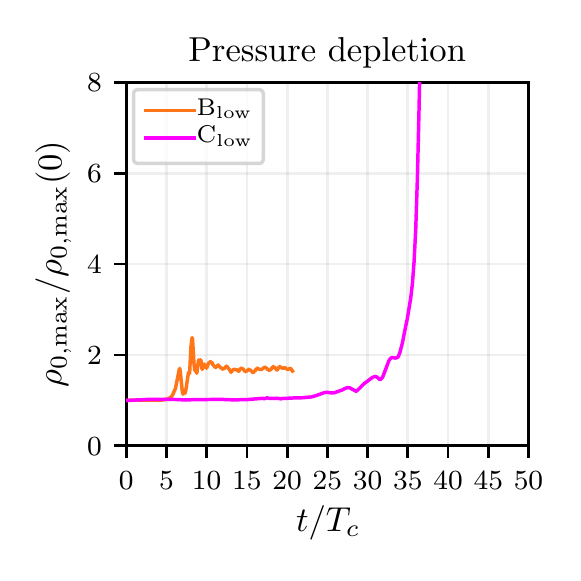} 
\includegraphics[width=5.9cm]{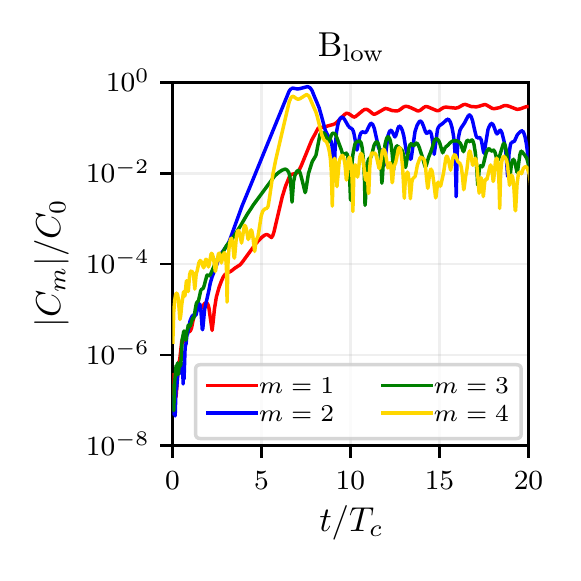} 
\includegraphics[width=5.9cm]{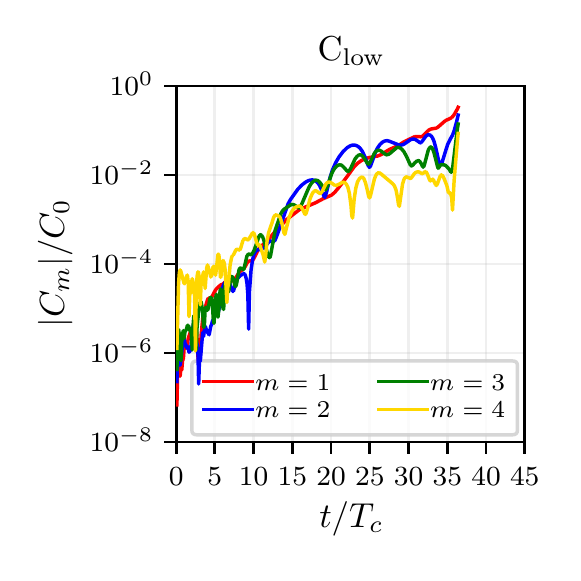}\\
\caption{Same as Fig.~\ref{fig:BClow_equi} but for the case of
  pressure depletion.  }
\label{fig:BClow_depl}
\end{figure*} 
Model $\rm B_{low}$, which is the lowest rest mass quasi-toroidal
model in our study, undergoes non-axisymmetric instabilities and
transitions dynamically from a quasi-toroidal shape to a spheroidal
one. This dynamical transition was observed for all types of
evolutions we considered. Here we describe the basic evolution of the
configuration and its unstable modes, and discuss the properties of
the final state in Sec.~\ref{subsec:final_state}.

In the left panel of Fig.~\ref{fig:BClow_equi} we plot $\rho_{0, \rm
  max}(t)$ for the equilibrium evolution of model $\rm B_{low}$. The
maximum rest mass density peaks at $\rho_{0, \rm max} \approx 2.5
\rho_{0,\rm max}(0)$ and subsequently oscillates around $\rho_{0,\rm
  max} \approx 1.7 \rho_{0,\rm max}(0)$ as the configuration evolves
towards a steady state. The fact that the maximum density does not
continue to increase demonstrates that this configuration is
dynamically stable against catastrophic collapse, but it will collapse
on secular timescales due to viscous/magnetic effects that redistribute
angular momentum, because the total rest mass exceeds the supramassive
limit rest mass~\cite{Shapiro:2000zh, Duez:2004nf, dlsss06a}. The
presence of the dip in the maximum density evolution after its first
peak is consistent with the feature discussed in \ref{subsec:Equil},
where the bar mode dominates the evolution. Thus, the maximum density
evolution alone suggests that the equilibrium evolution of model $\rm
B_{low}$ develops a bar-mode early on.  This can be seen in the
density mode decomposition, which is plotted in the center panel of
Fig.~\ref{fig:BClow_equi} and shows the $m=2$ mode dominance. The
plateau which the $m=2$ density mode exhibits in the time interval
$\sim 6-8T_c$, corresponds to the saturation of the bar mode. The time
of onset of the bar mode saturation coincides with the first peak of
the maximum rest mass density. Subsequently, a single, approximately
spheroidal, over-dense region forms giving rise to the second peak of
the maximum rest mass density (seen at $t\approx 8T_c$ in the left
panel of Fig.~\ref{fig:BClow_equi}). The formation of the single
spheroidal over-dense region signals the decay of the $m=2$ mode that
starts at $t\approx 8T_c$ (center panel of 
~Fig.\ref{fig:BClow_equi}).

Model $\rm C_{low}$ undergoes catastrophic collapse in the case of
equilibrium evolution, which is indicated by the rapidly 
increasing maximum density in the left panel of
Fig.~\ref{fig:BClow_equi}.  Model $\rm C_{low}$ is unstable against
the development of non-axisymmetric modes as shown in the right panel
of Fig.~\ref{fig:BClow_equi}, which drive the evolution toward
catastrophic collapse.

\subsubsection{Pressure depletion}

In Fig.~\ref{fig:BClow_depl} we show the evolution of the maximum
rest mass density, and the azimuthal density modes for the pressure
depletion perturbation. These are practically the same as the
equilibrium evolution for model $\rm B_{low}$, and very similar to the
equilibrium evolution for model $\rm C_{low}$. Only the
dominant non-axisymmetric density modes are slightly different,
but qualitatively the evolutions are very similar. The dynamical
transition to a spheroidal model for model $\rm B_{low}$, and the
collapse for model $\rm C_{low}$ in this case occurs practically on
the same timescale as in the equilibrium evolution case. This is a
clear indication that these quasi-toroidal models are not unstable
against quasi-radial perturbations. As in the equilibrium evolution
case, non-axisymmetric modes seeded at the level of truncation error
dominate the evolution.

\subsubsection{Non-axisymmetric perturbations}

To test whether the initial excitation of an $m=1$ or $m=2$ mode leads
to dominance of the excited non-axisymmetric mode, we now consider
$m=1$ and $m=2$ initial perturbations separately. Before discussing
the evolution of azimuthal density modes in each case, we first
discuss the general dynamics and how the evolution proceeds in the
$\rm B_{low}$ model for non-axisymmetric initial perturbations.  The
$\rm C_{low}$ model evolves similarly to the $\rm B_{low}$ model with
a strong one-arm mode developing in the case of an $m=1$ initial
perturbation, and a strong bar mode developing in the case of an $m=2$
initial perturbation, but ultimately leading to catastrophic collapse.

\begin{figure*}
\includegraphics[width=5.6cm]{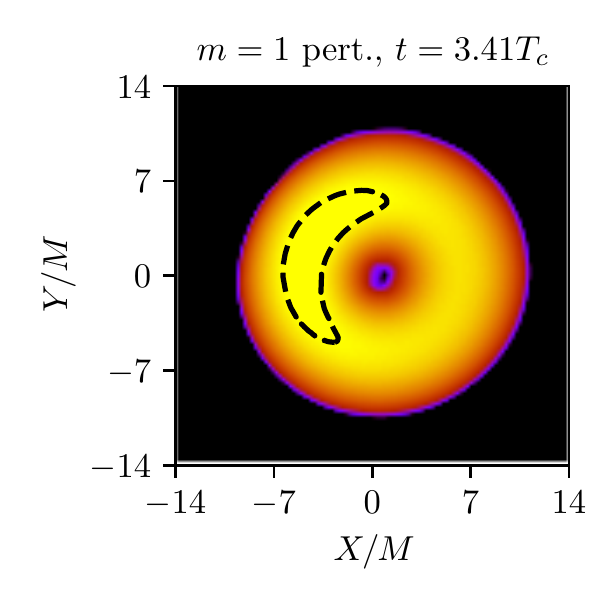}
\includegraphics[width=5.3cm]{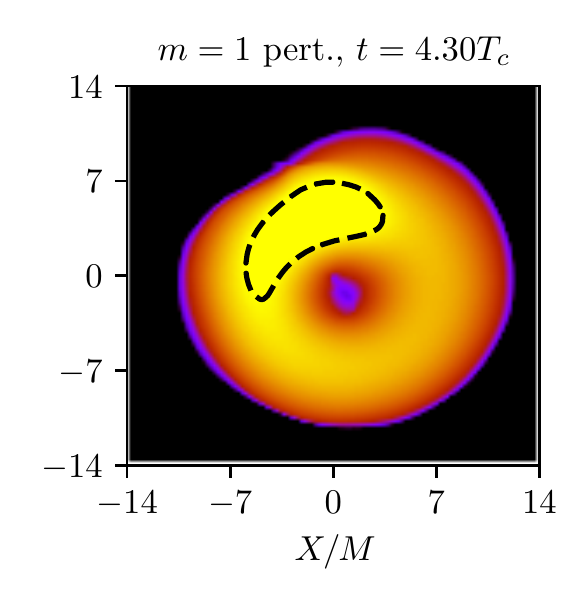}
\includegraphics[width=6.7cm]{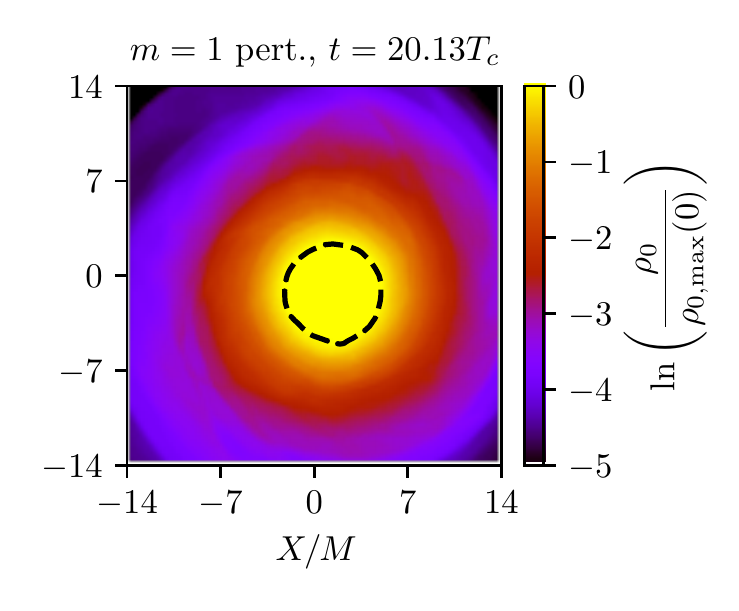}\\
\includegraphics[width=5.6cm]{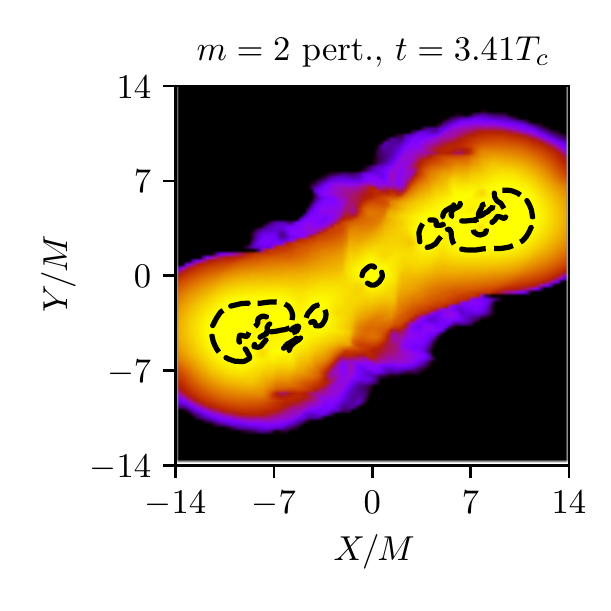}
\includegraphics[width=5.3cm]{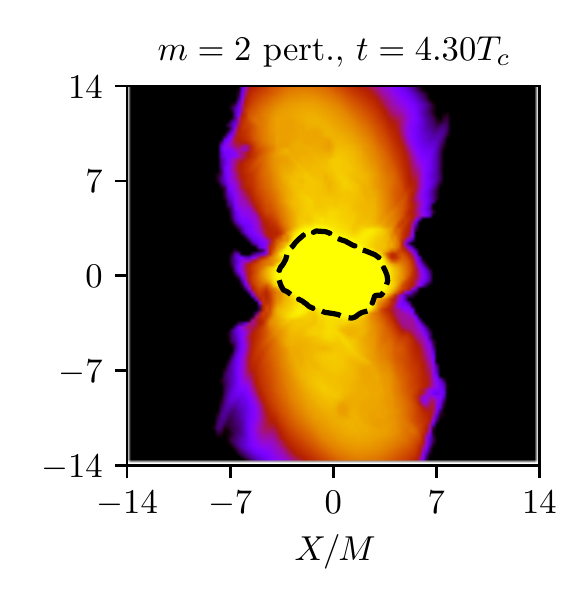}
\includegraphics[width=6.7cm]{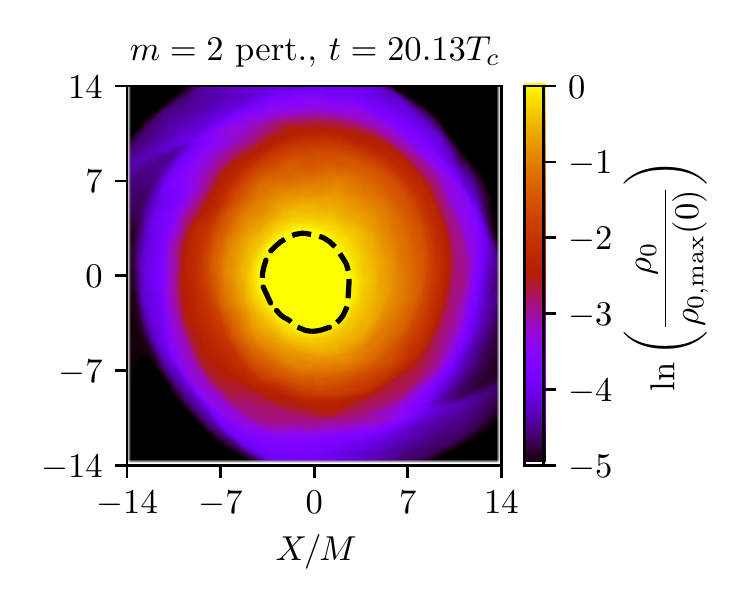}
\caption{Snapshots of equatorial rest mass density contours $\rho_0$,
  scaled to the maximum value at the start of simulation $ \rho_{0,\rm
    max}(0)$ for model $\rm B_{low}$ under non-axisymmetric initial
  perturbations.  The top (bottom) panels show the evolution under an
  initial $m=1$ ($m=2$) perturbation. The right panels show the final
  states in both cases. The dashed curves outline the boundary of the
  regions within which the rest mass density satisfies $\rho_0 \geq
  \rho_{0,\rm max}(0)$.}
\label{fig:Blo_snapshots}
\end{figure*}

\begin{figure*}
  \includegraphics[width=5.9cm]{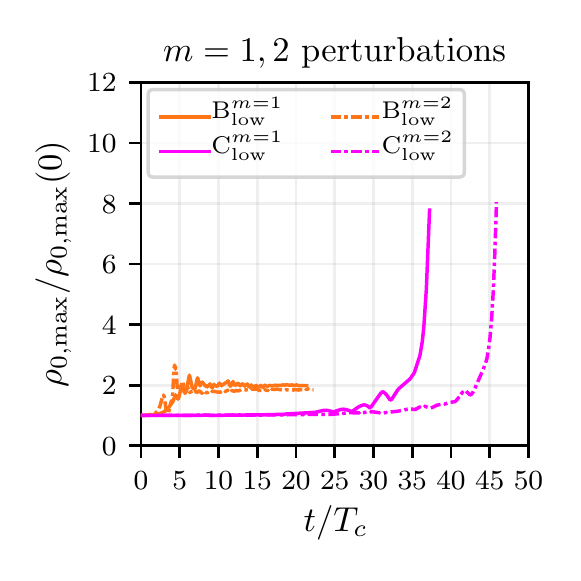}
\includegraphics[width=5.9cm]{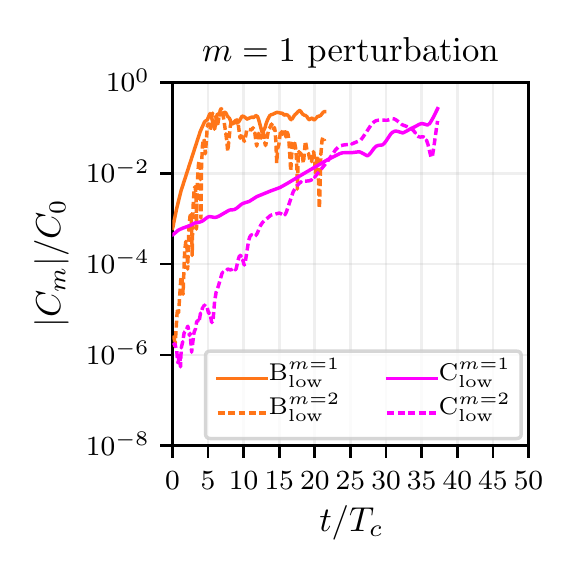} 
\includegraphics[width=5.9cm]{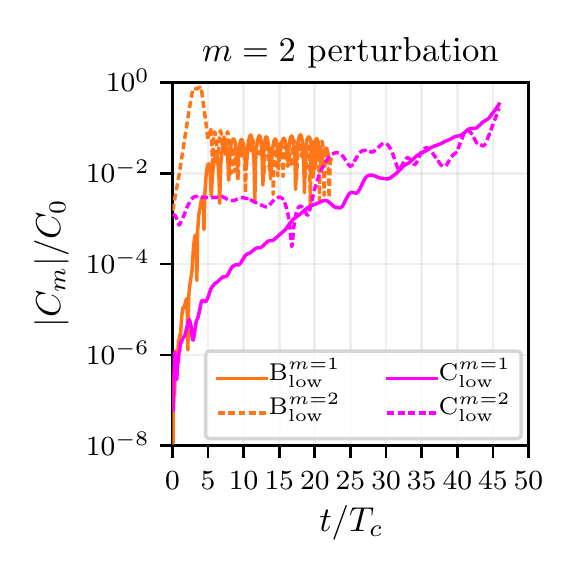} 
\caption{Left panel: Maximum rest mass density as function of time for
  the $\rm B_{low}$ and $\rm C_{low}$ models in the case of
  non-axisymmetric $m=1$ (solid lines) and $m=2$ (dash-dotted lines)
  rest mass density initial perturbations. The orange and magenta lines
  correspond to the $\rm B_{low}$ and $\rm C_{low}$ models,
  respectively.  Center panel: evolution of the dominant $m=1$ 
  (solid lines)
  and $m=2$ (dashed lines) non-axisymmetric density modes for the $\rm
  B_{low}$ (orange lines) and $\rm C_{low}$ models (magenta lines) in
  the case of an $m=1$ initial perturbation.  Right panel: Same as 
  the center
  panel but for an $m=2$ initial perturbation.}
\label{fig:BClow_m1m2}
\end{figure*}

In Fig.~\ref{fig:Blo_snapshots} we show equatorial snapshots of the
rest mass density for the $\rm B_{low}$ model in the cases of $m=1$
and $m=2$ perturbations. In the plots we also indicate with dashed
lines the regions where $\rho_0 \geq \rho_{0,\rm max}(0)$, which early
on show where in the star the one-arm or bar modes begin to grow.

In the case of the $m=1$ initial perturbation, first a single
over-dense region develops in the high-density ring around the center
of mass (see the top left panel of Fig.~\ref{fig:Blo_snapshots}). The
growth of the $m=1$ mode occurs on a dynamical timescale and the
over-dense region quickly grows (top center panel of
Fig.~\ref{fig:Blo_snapshots}), eventually migrating toward the
geometric center of the original configuration, and settling there
(top right panel of Fig.~\ref{fig:Blo_snapshots}).

When an $m=2$ perturbation is initially excited, the $m=2$ mode is
seen to dominate throughout the evolution. A bar develops early on,
and the two arms of the bar continue to separate into a dumbbell like
configuration, with two over-dense regions momentarily orbiting around
a third over-dense region near the center of mass (see the bottom left
panel of Fig.~\ref{fig:Blo_snapshots}). Subsequently, the bar mode
saturates, after which the two over-dense arms coalesce with the
central over-dense region to form a \emph{single} over-dense core (see
the bottom center panel of Fig.~\ref{fig:Blo_snapshots}). The
configuration eventually settles toward a spheroidal shape (bottom right
panel of Fig.~\ref{fig:Blo_snapshots}).

\begin{figure*}
\includegraphics[width=8.5cm]{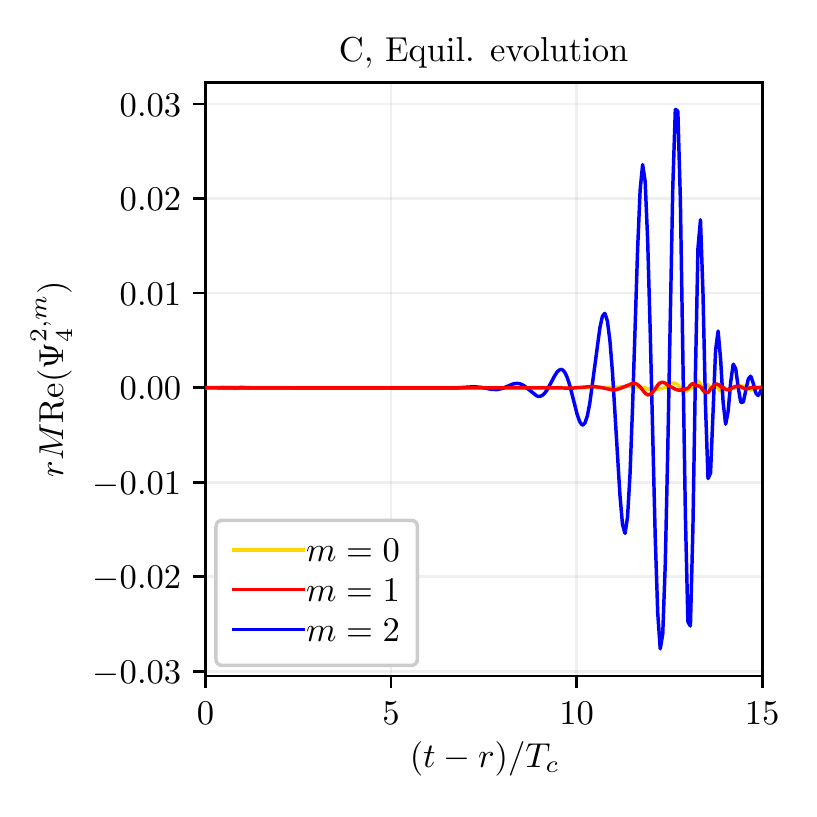}
\includegraphics[width=8.3cm]{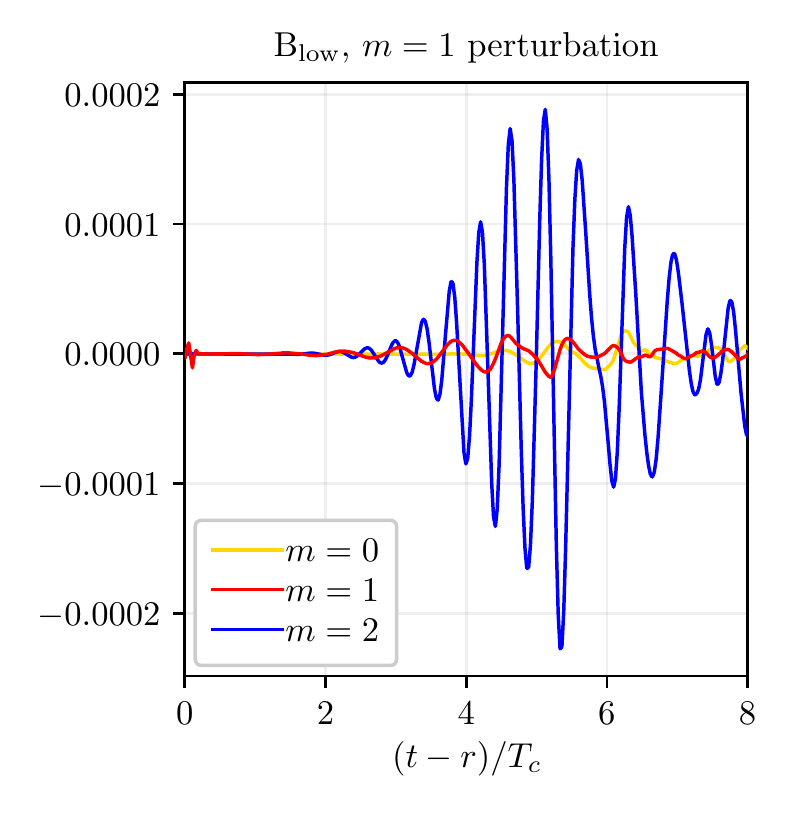}
\caption{The real part of the Newman-Penrose scalar $s=-2$
  spin-weighted spherical harmonic modes $\Psi_4^{2,m}$ (multiplied by
  the coordinate radius of the extraction spherical surface $r$ and
  the ADM mass $M$) as a function of $t-r$ (scaled by the central
  period $T_c$) for two representative evolutions presented in this
  work. We focus on the contribution from the dominant $l=2$, $m=0$
  (yellow lines), $m=1$ (red lines) and $m=2$ (blue lines) modes.  The
  model and type of perturbation are shown at the top of each panel.
}
\label{fig:psi4_ex}
\end{figure*}
In Fig.~\ref{fig:BClow_m1m2}, we show the evolution of the maximum
rest mass density and the density mode decomposition for the $\rm
B_{low}$ and $\rm C_{low}$ models under initial non-axisymmetric rest
mass density perturbations. For the $\rm B_{low}$ model, at late times
$\rho_{0,\rm max}$ exhibits small oscillations around the value
$\rho_{0, \rm max} \approx 1.85\rho_{0,\rm max}(0)$ for the $m=1$
initial perturbation case, and at $\rho_{0, \rm max} \approx
2\rho_{0,\rm max}(0)$ in the $m=2$ initial perturbation case. Note
that in the cases of equilibrium evolution and pressure depletion the
quasi-steady state maximum rest mass density is closer to
$1.7\rho_{0,\rm max}(0)$. This result suggests that the remnants may
be settling to similar (though not identical) final configurations. We
investigate this issue further in the next Section. The center and
right panels of Fig.~\ref{fig:BClow_m1m2} show only the dominant $m=1$
and $m=2$ density mode amplitudes. The plots demonstrate clearly that
when an $m=1$ ($m=2$) mode is initially excited, then the $m=1$
($m=2$) density mode dominates the evolution.

Model $\rm C_{low}$ collapses to form a BH earlier in the case of an $m=1$
initial perturbation than for an $m=2$ initial perturbation. This
suggests that the one-arm mode may be growing on a faster timescale
than the bar mode. However, the fact that in the $m=1$ initial
perturbation the collapse does not occur much earlier than in the
equilibrium or pressure depletion cases suggests that the mode we
excite seeds the unstable mode eigenfunction, but may not be the true
eigenfunction. Nevertheless, the general result is consistent with our
findings for the high rest mass B and C models (see the left panel of
Fig.~\ref{fig:m1m2_densmode}), i.e., the non-axisymmetric mode that is
excited first dominates the subsequent evolution.

Regardless of the details, our results demonstrate the importance of
$m=1$ modes, as has already been pointed out in the studies of NS
mergers in \cite{Paschalidis:2015mla}, where a one-arm instability
develops in long-lived remnants. In addition, our calculations
demonstrate the significance of an $m=1$ mode in triggering
catastrophic collapse.  Therefore, imposing $\pi$-symmetry in BNS
merger calculations, which is often employed to save computational
resources (e.g.~\cite{Baiotti:2006wn, Manca:2007ca,
  Rezzolla:2010fd,Giacomazzo:2010bx,Zhang:2017fsy}), should be
avoided.

\subsection{Gauge-invariant measure of non-axisymmetric mode development}

As discussed in Sec.~\ref{subsec:diagnostics}, the density mode
decomposition using Eq.~\eqref{eq:densmode_decomp} is a
gauge-dependent diagnostic of the dominant modes that develop during
the evolution.  To ensure the features of evolution discussed thus far
are not gauge artifacts, we also study the gravitational wave
signatures using the Newman-Penrose formalism. We focus on two
representative cases that include one massive model that collapses to
a BH (the C model) and our low-mass model the ($\rm B_{low}$) that 
does not undergo collapse. In particular, the evolutions we consider 
here are the C model under no explicit initial perturbations, and 
$\rm B_{low}$ under an $m=1$ initial perturbation.

In Fig.~\ref{fig:psi4_ex}, we show the real part of the $s=-2$
spin-weighted spherical harmonic modes $\Psi_4^{l,m}$. In general, the
same conclusions reached by studying the density modes $C_m$ may be
reached if we consider $\Psi_4^{2,m}$ as a measure of non-axisymmetric
mode growth. We find that quasi-toroidal stars exhibit growth of the
$m=1$ and $m=2$ GW modes, even in cases where neither of these modes
were explicitly excited.  However, the relative amplitude of the GW
modes is not the same as in the density mode decomposition. For
example, even when an $m=1$ mode is initially excited, the GW $m=1$
mode does not dominate over the $m=2$ mode (as shown in the right
panel of Fig.~\ref{fig:psi4_ex}). Nevertheless, in all cases
considered, we observe strong growth of quasi-radial and
non-axisymmetric modes consistent with the results presented in
Secs. \ref{subsec:resABC} and \ref{subsec:Blow_res}.

\section{Discussion}
\label{sec:discussion}
In this Section, we further discuss the results of our simulations and
compare them with the results in the literature on differentially
rotating $\Gamma=2$ polytropes. We focus on the role that different
properties may play in the evolution of our initial data. We also
discuss further the final state that model $\rm B_{low}$ reaches after
it settles, and the implications of our findings on cosmic censorship
and the fragmentation instability of quasi-toroids.

\subsection{Role of $T/|W|$}
\label{subsec:T_W}

Generally we find that our massive quasi-toroidal models with large
values of $T/|W|$ collapse on short timescales due to the growth of
non-axisymmetric modes. It is possible that for a given degree of
differential rotation there exists a critical value of $T/|W|$ which
signals the onset of instability to non-axisymmetric modes for
quasi-toroidal neutron stars described by a $\Gamma=2$ polytropic
EOS. In \cite{Loffler2015}, bounds were placed on the critical value
$T/|W|_{\rm crit}$ which indicates dynamical instability to the
growth of the bar mode for $\Gamma=2$ quasi-toroids when $\hat
A^{-1}=1$ and for masses $1M_\odot-2.5M_\odot$ assuming a polytropic
constant $\kappa=165M_\odot^2$. The following fit for $T/|W|_{\rm crit}$ as a
function of the rest mass was derived for $\Gamma=2$ quasi-toroidal
models,
\begin{eqnarray}
\label{eq:Loffler_bound}
T/|W|_{\rm crit}(M_0) &  = & 0.2636 - 0.0047 \dfrac{M_0}{M_\odot} \nonumber \\
                   & = &  0.2636 - 0.0108 \dfrac{M_0}{M_{0,\rm max}^{\rm TOV}} 
\end{eqnarray}
such that all $\Gamma=2$ quasi-toroids of rest mass $M_0$ with $T/|W|
< T/|W|_{\rm crit}$ should be dynamically stable against the growth of
the bar mode instability.  We can now test the applicability of Eq.
\eqref{eq:Loffler_bound} to our models. We convert from polytropic
units to units of $M_\odot$ by using the value of the polytropic
constant adopted in \cite{Loffler2015}.  The $\rm C_{low}$ model
considered in this work is the model with the lowest value of $T/|W|$
which is unstable to the growth of a dynamical bar mode. Inserting the
rest mass $M_{0}^{\rm C_{low}}\simeq 4.18M_\odot$ of the $\rm C_{low}$
model into Eq.~\eqref{eq:Loffler_bound}, we find a critical value of
$T/|W|^{\rm C_{low}}_{\rm crit}=0.244$, suggesting that all $\Gamma=2$
quasi-toroidal stars of rest mass equal to that of the $\rm C_{low}$
model and with $T/|W| < T/|W|^{\rm C_{low}}_{\rm crit}$ should be
stable against the growth of a dynamical bar mode. We find that the
$\rm C_{low}$ model slightly violates this bound, as it has
$T/|W|=0.238 \approx 0.98$ $T/|W|^{\rm C_{low}}_{\rm crit}$ and is
still unstable to the growth of a dynamical bar mode. However, the
violation is not significant and we cannot conclude that
Eq.~\eqref{eq:Loffler_bound} is inconsistent with our
findings. Further studies are necessary to probe the applicability of
Eq.~\eqref{eq:Loffler_bound}.
\begin{figure*}
\includegraphics[width=4.8cm]{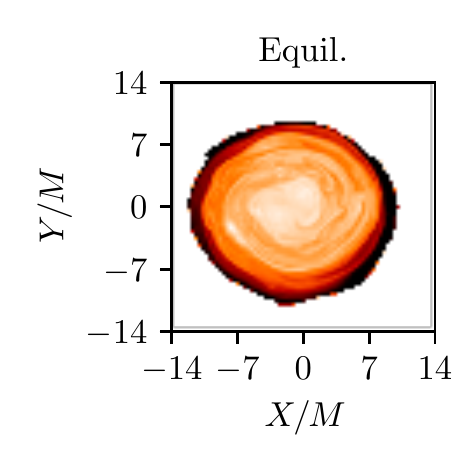} 
\includegraphics[width=3.8cm]{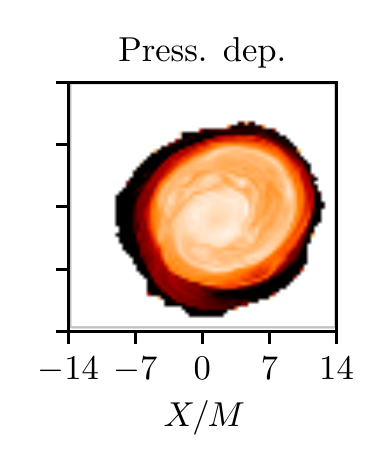}
\includegraphics[width=3.8cm]{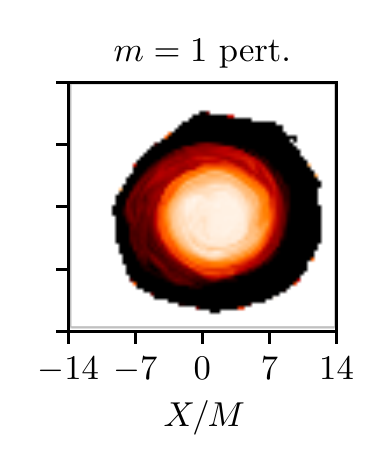} 
\includegraphics[width=5.2cm]{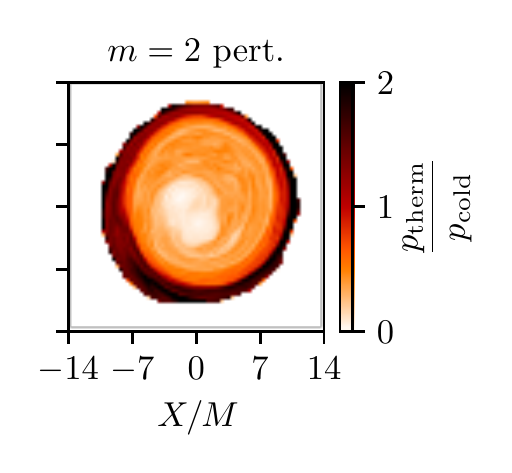}
\caption{Contour of the ratio of thermal pressure to cold pressure on
  the equatorial plane corresponding to the final state of the $\rm
  B_{low}$ model under all initial perturbations.  Only densities with
  $\rho_0 \geq 10^{-3}\rho_{0,\rm max}$ are shown, where $\rho_{0,\rm
    max}$ corresponds to the maximum rest mass density at the time of
  the snapshots. All snapshots correspond to $t=20.13T_c$; the second
  from right and rightmost snapshots corresponds to the top and bottom
  right panels of Fig.~\ref{fig:Blo_snapshots} for the $m=1$ and $m=2$
  perturbations, respectively.}
\label{fig:P_th}
\end{figure*}
As discussed in Sec.~\ref{sec:res}, all quasi-toroidal models
considered in this work were unstable to non-axisymmetric mode
growth. Considering a similar analysis to that provided above for the
$\rm C_{low}$ model, we find that all quasi-toroidal models respect
the bound on $T/|W|_{\rm crit}$ set by Eq.
\eqref{eq:Loffler_bound}. This suggests that the applicability of
Eq.~\eqref{eq:Loffler_bound} may be extended to $\hat A^{-1}\neq 1$
and well outside the mass range studied in~\cite{Loffler2015}. 

However, our quasi-toroidal models are unstable not only to the bar
mode, but also to the one-arm mode. A similar study to that presented
in \cite{Loffler2015} and a derivation of a formula similar to
Eq.~\eqref{eq:Loffler_bound} but with a focus on the growth of the
one-arm mode is needed to determine the stability of quasi-toroidal
$\Gamma=2$ configurations against the growth of the $m=1$ mode, but is
outside of the scope of this work. We also point out that our model A
has $T/|W|\sim 0.3$ and does not develop any non-axisymmetric
modes. Therefore, our study demonstrates that $T/|W|$ alone does not
determine the type of instability in a differentially rotating
configuration. This is consistent with the existence of the
low-$T/|W|$ instability.

\subsection{Role of the rest mass}

In our simulations, the value of $M_0$ appears to control the final
state of the configuration, i.e., whether the configuration collapses
to a black hole on a dynamical time or not.  Coincidentally, all of 
our models, except A and $\rm B_{low}$, have rest masses which well 
exceed the threshold mass for prompt collapse found in BNS merger 
simulations of
$\Gamma=2$ polytropes \cite{Shibata:2003ga} ($M_{0}^{\rm C_{low}} >
1.75M_{0,\rm max}^{\rm TOV}$), and all of these models undergo
collapse to BH on dynamical timescales. By contrast, the 
$\rm B_{low}$ model does not collapse to a BH by the
end of the simulation and shows no signs that collapse will
ensue. This model's rest mass $M_{0}^{\rm B_{low}} =1.36M_{0,\rm
  max}^{\rm TOV}$ is lower than the lower-bound on the threshold rest
mass $1.65M_{0,\rm max}^{\rm TOV}$ for prompt collapse to a BH
\cite{Shibata:2003ga}.

Although the above discussion suggests that even in isolated rotating
neutron star models, the total rest mass controls whether there will
be collapse to a BH on dynamical timescales or not, we point out that
the dimensionless angular momentum of model $\rm B_{low}$ is higher
than all other cases and larger than in BNS mergers. This excess
angular momentum may provide additional centrifugal support against
collapse. More models need to be considered for a complete study, and
to test whether the threshold mass for prompt collapse found in BNS
mergers also provides the line of stability against collapse in
isolated neutron stars that may model BNS merger remnants. This is
related to the study of~\cite{Bauswein:2017aur} who derived the
threshold-mass for prompt collapse by use of rotating {\it spheroidal}
neutron star models.

\subsection{Final state of the $\rm B_{low}$ model}
\label{subsec:final_state}
\begin{figure}
\includegraphics[width=8.5cm]{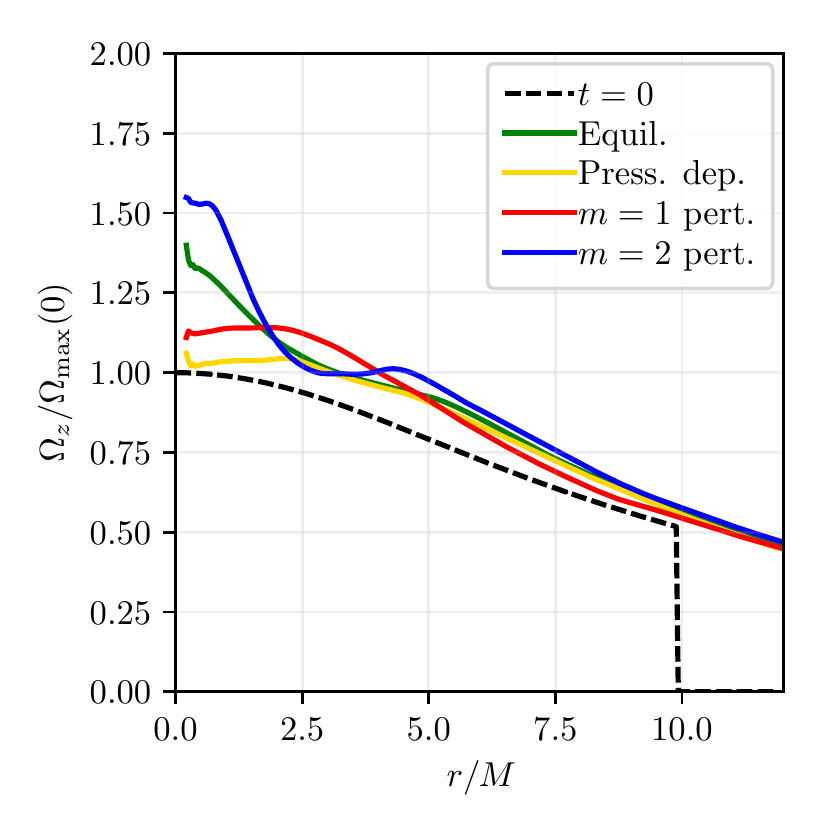}   
\caption{Azimuthally- and time-average angular velocity (normalized by
  the initial maximum angular velocity) vs distance on the equatorial
  plane in the case of equilibrium evolution (green line), pressure
  depletion (yellow line), $m=1$ initial perturbation (red line) and
  $m=2$ initial perturbation (blue line) near the end of the
  simulations. The dashed black line corresponds to the initial
  angular velocity profile.  }
\label{fig:angvel_post}
\end{figure}

Given that model $\rm B_{low}$ does not collapse to a BH, an
interesting question is whether this model evolves to the same final,
dynamically stable configuration under all evolutions considered. 
Moreover, given the presence of shocks during the
evolutions it is also interesting to test whether significant thermal
pressure support exists. In this section we investigate these
questions.

Shocks arise during the evolution, and the amount of shock heating
differs between the one-arm mode and bar mode evolutions. To test for
the impact of shock heating we exploit the fact that the total fluid 
pressure $p$ can be expressed as a sum of the cold pressure $p_{\rm
  cold}$ and thermal pressure $p_{\rm therm}$,
\begin{equation}\label{eq:ptot}
p = p_{\rm cold} + p_{\rm therm},
\end{equation}
where the cold part of the EOS is described by
Eq.~\eqref{eq:poly}. Our initial models are cold, i.e., $p_{\rm
  therm}=0$ at $t=0$. As the evolution proceeds, shock heating can
take place and $p_{\rm therm}$ grows.  The separation in Eq.
\eqref{eq:ptot} allows us to determine the contribution from the
thermal pressure in the final configuration as follows,
\begin{equation} \label{eq:ptherm_pcold}
\dfrac{p_{\rm therm}}{p_{\rm cold}} = \dfrac{p}{\kappa\rho_0^2} - 1.
\end{equation}
If the thermal pressure is a significant component of the total
pressure, it may be that the final configuration is only possible
because of additional thermal support. This would imply that if the
configuration were allowed to cool, it might collapse to a
BH~\cite{Paschalidis:2012ff}. In Fig.~\ref{fig:P_th} we show snapshots
of equatorial contours of $p_{\rm therm}/p_{\rm cold}$ after the
configurations in the different $\rm B_{low}$ evolutions have settled
down to an approximately steady state at $t=20.13T_c$. We focus on
regions where the density is $\rho_0 \geq 10^{-3} \rho_{0, \rm max}$,
where the bulk of the matter is. We find similar results under all
evolutions, where thermal support near the end of the simulation is
small in the regions near the core. The thermal pressure can be
significant slightly outside the core and in regions further
out. Therefore, the remnants are primarily cold, but have experienced
different amounts of shock heating. Cases with a dominant $m=2$ mode
during the early stages of the evolution (i.e., the equilibrium
evolution, pressure depletion, and $m=2$ initial perturbation) look
more similar amongst themselves, and distinct from the case where an
$m=1$ initial perturbation is excited.

Figure~\ref{fig:angvel_post} shows the time- and azimuthally-averaged
radial rotation profiles, i.e, angular velocity
($\Omega_z=u^\phi/u^t$) in the $z$ direction vs distance on the
equatorial plane, that correspond to the final states of model $\rm
B_{low}$ shown in Fig \ref{fig:P_th}. Also shown is the initial
angular velocity profile. The plot is only meant to be illustrative,
because the angular velocity we compute is not gauge invariant unlike
the initial angular velocity. We average the rotational profiles over
the azimuthal direction, while accounting for the shift of the
configuration's center of rest mass from the initial center of mass,
i.e., the coordinate origin. We also average the rotational profiles
over a time window of $\Delta t\simeq T_c$.  The rotation profiles of
the final state of the $\rm B_{low}$ model are distinct from each
other. All final configurations are highly differentially
rotating. Steep gradients are seen near the center of the
configurations for the equilibrium evolution and $m=2$ perturbation
cases, corresponding to high differential rotation in the innermost
region near the core. The central regions in the cases of pressure
depletion and for an $m=1$ perturbation show profiles with relatively
lower amounts of differential rotation, and the angular velocity
instead increases away from the core. The central region is surrounded
by a distribution of matter with decreasing angular velocity for
$r\gtrsim 5 M$ in all cases. Taking the final angular velocity
profiles at face value, it does not appear possible to approximate the
angular velocity of these configurations by the one parameter KEH
rotation law in Eq.~\eqref{eq:komatsu_rotlaw}, due to their
non-monotonic nature. As such, we do not try to approximate the final
states of the $\rm B_{low}$ evolutions with stars described by the KEH
law. We conclude that the final states reached in different evolutions
of the $\rm B_{low}$ model are close to each other but not identical,
as was already suggested by the different value of maximum rest mass
density to which they settle.

\subsection{Fragmentation instability  and cosmic censorship}

Our initial data range over values of the dimensionless spin parameter
$J/M^2$ which are both less than and greater than the Kerr bound. If
cosmic censorship holds \cite{Penrose:1969pc}, models with $J/M^2 > 1$
will not be able to form a BH unless they shed or redistribute angular
momentum through mass ejection, gravitational waves or other
mechanisms \cite{Giacomazzo:2011cv} or possibly fragment
\cite{Zink:2006qa}.

In all collapsing cases, we were only able to locate a single AH, and
the evolution of the lapse function is consistent with a single BH
forming.  Therefore, we do not find that our stars could collapse to
form binary black holes as in \cite{Reisswig:2013sqa}, even in the
most massive cases.

In \cite{Giacomazzo:2011cv}, dynamically unstable, differentially
rotating, supra-Kerr models of $\Gamma=2$ neutron stars could not be
found and supra-Kerr models could only be induced to collapse through
severe pressure depletion. In this work we found that dynamically
unstable supra-Kerr models do exist for $\Gamma=2$ polytropes (for
instance, both the B and C models collapsed to a BH on dynamical
timescales even in the case of equilibrium evolution). None of the
dynamically unstable supra-Kerr models were found to produce naked
singularities. In all cases where the resulting BHs were evolved long
enough to settle into approximately steady states, we find that they
are surrounded by disks with rest masses as large as $18\%$ the rest
mass of the initial configuration. The BHs that form in all cases that
undergo catastrophic collapse (initially either sub- or supra-Kerr)
have dimensionless spin parameter $a\simeq 0.85$, and hence not close
to unity. We have checked that GWs carry away $O(1\%)$ of the initial
angular momentum, and that total angular momentum is conserved to
within 1\%. The above imply that the remnant disks carry a significant
amount of the initial angular momentum. These simulations provide yet
another example in which cosmic censorship is respected.

\section{Conclusions}
\label{sec:conclusion}

In this work we performed dynamical simulations in full general
relativity to investigate the stability of differentially rotating,
high mass spheroidal and quasi-toroidal $\Gamma=2$ polytropic models
of neutron stars. Compared to previous works studying the stability of
differentially rotating hypermassive neutron stars, our work probes a
part of the parameter space that has not been probed before, namely
the part corresponding to highly quasi-toroidal, and very massive
stars (as massive as $\sim 4$ times the TOV limit mass). Recent work,
which discovered these extreme configurations, suggested that massive
quasi-toroidal configurations could have important consequences for
neutron star mergers or core collapse supernovae. Indeed, the
existence of such massive equilibria might suggest that much more
massive remnants than previously found could exist in these
astrophysical scenaria. But, for this to be the case, such extreme
quasi-toroidal configurations would have to be dynamically
stable. Thus, here we initiated a study of the dynamical stability of
these extreme equilibria.

Four of the five initial equilibria we investigated are
quasi-toroidal: models B and C, which are the most massive, and models
$\rm B_{low}$ and $\rm C_{low}$, with masses closer to remnants that
could form following a BNS merger. The fifth initial configuration we
considered is the most massive (type A) spheroidal star, also with
astrophysically relevant rest mass. Apart from model $\rm B_{low}$, we
found that all models underwent catastrophic collapse to single BHs
under various types of initial perturbations or no perturbations at
all.  The most massive spheroidal model was unstable only against a
quasi-radial perturbation. By contrast, all quasi-toroidal
configurations were unstable to the development of non-axisymmetric
instabilities.  We found that the dominant non-axisymmetric modes are
either the $m=1$ or $m=2$ modes, which grow on very similar
timescales. Our simulations indicate that the first non-axisymmetric
mode to be seeded early on in the evolution of a quasi-toroidal star
is the mode that dominates the evolution. Thus, when the $m=1$ mode is
excited a one-arm instability takes over, whereas when the $m=2$ mode
is excited a bar-mode instability dominates. We find that in some
cases the one-arm mode may dominate over the bar mode, when no
explicit perturbations are seeded initially, but perturbations are
always excited at the level of truncation error in our simulations
because our grid coordinates are slightly shifted to avoid the origin.
Our findings further demonstrate the importance of the $m=1$ mode for
the stability of differentially rotating neutron stars, that was
recently pointed out
in~\cite{Paschalidis:2015mla,PEFS2016,East:2016zvv}.

 \begin{figure*}[t]
\includegraphics[width=5.9cm]{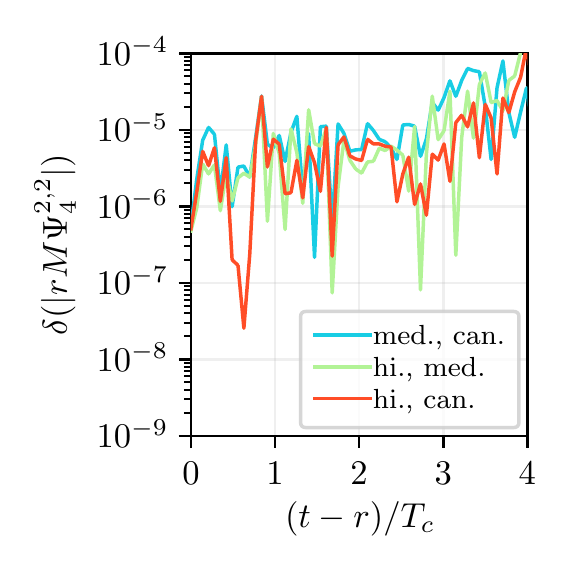} 
\includegraphics[width=5.9cm]{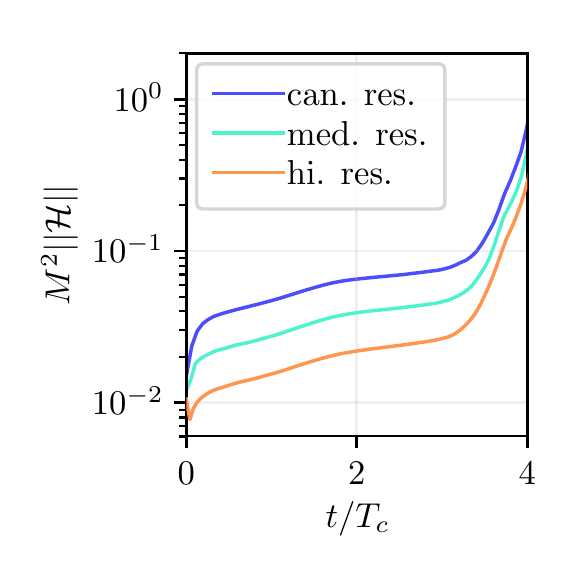} 
\includegraphics[width=5.9cm]{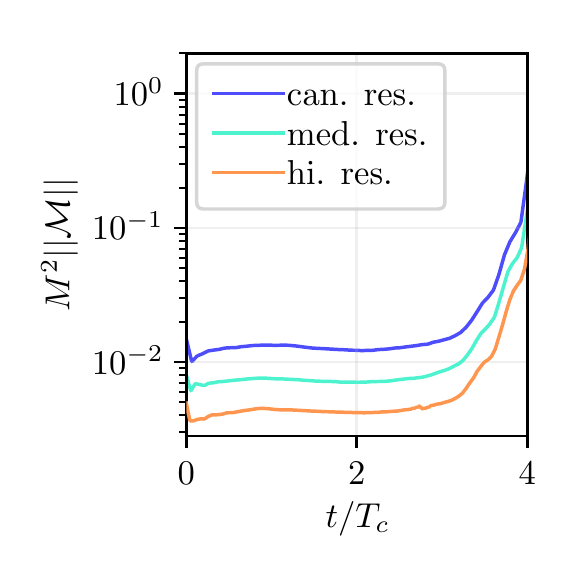} 
\caption{ Left panel: Convergence of $|\Psi_4^{2,2}|$ as a function 
of $(t-r)/T_c$ in the case of the B model under an $m=2$ 
 perturbation. The 
  left panel shows the difference of $|\Psi_4^{2,2}|$ between the 
  medium and canonical resolution runs (cyan line, labeled ``med., 
  can.''), the medium and high resolution runs (green
  line, labeled ``hi., med."), and the high and canonical resolution 
  runs (orange line, labeled ``hi., can."), scaled assuming second 
  order convergence. Center panel: L2 norm of the Hamiltonian 
  constraint times the squared ADM mass $M^2||\mathcal{H}||$ as a 
  function of time for the B model under an $m=2$ perturbation. Shown 
  are the canonical (blue line), medium (cyan line), and high (orange 
  line) resolution runs.  Right panel: Same as the center panel, but 
  for the L2 norm of the momentum constraint $||\mathcal{M}||$.}
\label{fig:convergence}
\end{figure*}

We considered the role that the ratio of rotational kinetic energy to
gravitational binding energy $T/|W|$ played in the stability of these
equilibria. We find that the prediction of \cite{Loffler2015} (which
applies to a particular degree of differential rotation and a certain
range of masses) for the critical value $T/|W|_{\rm crit}$ marking 
the onset of the dynamical bar mode instability is largely consistent 
with the models studied here. Our model $\rm C_{low}$, with the 
lowest
value of $T/|W|$, slightly violates the prediction but we cannot
conclude that it is inconsistent with it. However, we study models 
with $T/|W|>T/|W|_{\rm crit}$ that do not undergo the bar mode 
instability, indicating that there are other important parameters at 
play for a configuration to be unstable to the bar mode instability. 
We also considered the role of the rest mass in determining the final 
state of quasi-toroidal configurations. Our study suggests that the 
total rest mass appears to determine primarily whether collapse to BH 
will ensue or not. However, more detailed studies are necessary to 
solidify these conclusions.

Our lowest-mass quasi-toroidal model ($\rm B_{low}$) underwent a
transition to a spheroidal solution either through a bar mode or a
one-arm mode instability, depending on the mode that is excited
first. However, we did not excite higher non-axisymmetric modes, and
it is conceivable that all quasi-toroidal models we considered are
unstable against $m>2$ modes, too \cite{Friedman:1978hf}. We find that
the dynamically found, highly differentially rotating,
quasi-stationary, spheroidal solutions resulting from the evolution of
model $\rm B_{low}$ are similar but distinct from each other. These
final configurations have a small thermal pressure component at their
cores, but thermal pressure is non-negligible far from the core. We
also found that their angular velocity profiles do not appear to be
reasonably approximated by the KEH rotation law. All of the final
states of the $\rm B_{low}$ model are dynamically stable, but
secularly unstable due to dissipative effects.

We investigated the properties of the BHs formed in the collapsing
models, and found that all resulting BHs have high dimensionless spin
(about 0.85) by the end of simulation and cosmic censorship is always
respected even when the initial solutions exceed the Kerr limit. We
did not find evidence of the formation of multiple BHs.  Our study
explicitly shows that exceeding the Kerr bound initially does not
imply the dynamical stability of a rotating stellar configuration.

Our work shows that the existence of extreme quasi-toroidal neutron
star equilibrium solutions, which support a mass well exceeding the BNS
threshold mass for prompt collapse to a BH, does not imply that BNS
merger remnants can be very massive, too. Moreover, highly
quasi-toroidal models of neutron stars appear to be dynamically
unstable against the development of non-axisymmetric instabilities,
and will either collapse to a BH or transition to a dynamically 
stable, spheroidal, differentially rotating configuration.

A few caveats for the present work are in order. First, we did not
scan the entire solution space of differentially rotating
quasi-toroidal solutions, nor did we build our initial models to
correspond to a particular sequence (i.e., constant rest mass or
constant angular momentum sequences). The rationale in this work was to
probe the most massive, differentially rotating configurations recently
found in the literature, and include a few lower mass models of the
quasi-toroidal type to test if configurations more massive than what
is achievable in BNS mergers can be dynamically stable. More realistic
descriptions of the matter may play an important role in the evolution
of quasi-toroidal models, but we do not expect it to change our basic
conclusion that massive, quasi-toroidal models of neutron stars built
with the KEH differential rotation law are generically dynamically
unstable. In particular, analogous models to those studied here
described by the KEH rotation law have been shown to exist for
realistic, hybrid hadron-quark, and strange quark matter equations of
state \cite{Espino:2019ebx, Bozzola:2019tit, Szkudlared2019}. The KEH
law may not suitably describe the remnants of BNS mergers
\cite{Paschalidis:2015mla, Abdikamalov:2008df, Galeazzi:2011nn}. More
realistic rotation laws \cite{Uryu:2017obi} could possibly lead to
different stability properties of quasi-toroidal stars. Finally, the
effects of magnetic fields could significantly affect the evolution of
quasi-toroids.  Magnetic braking and turbulent magnetic viscosity may
act to remove differential rotation on short timescales
\cite{Shapiro:2000zh, Duez:2004nf, dlsss06a, Ruiz:2019ezy}, leading to
faster collapse in the types of stars studied here. We leave a more
systematic investigation of all these topics for future work.

\section*{Acknowledgments}

PE and VP would like to thank KITP for hospitality during the
GRAVAST19 program, where part of this work was completed. This
research was supported in part by National Science Foundation (NSF)
Grant PHY-1662211, and NASA Grant 80NSSC17K0070 at the University of
Illinois at Urbana-Champaign, NSF grant PHY-1707526 to Bowdoin
College, and through sabbatical support from the Simons Foundation
(Grant No. 561147 to TWB). The simulations presented in this work were
carried out in part on the Ocelote and ElGato clusters at the
University of Arizona, the Blue Waters supercomputer at NCSA, and the
Stampede2 cluster at TACC under XSEDE allocation PHY180044. Research
at KITP is supported in part by the National Science Foundation under
Grant No. NSF PHY-1748958. The Blue Waters sustained-petascale
computing project is supported by the National Science Foundation
(awards OCI-0725070 and ACI-1238993) and the State of Illinois. Blue
Waters is a joint effort of the University of Illinois at
Urbana-Champaign and its National Center for Supercomputing
Applications.

\appendix

\section{Resolution study and grid effects}
\label{app:high_res}

\begin{figure}
\includegraphics[width=7.5cm]{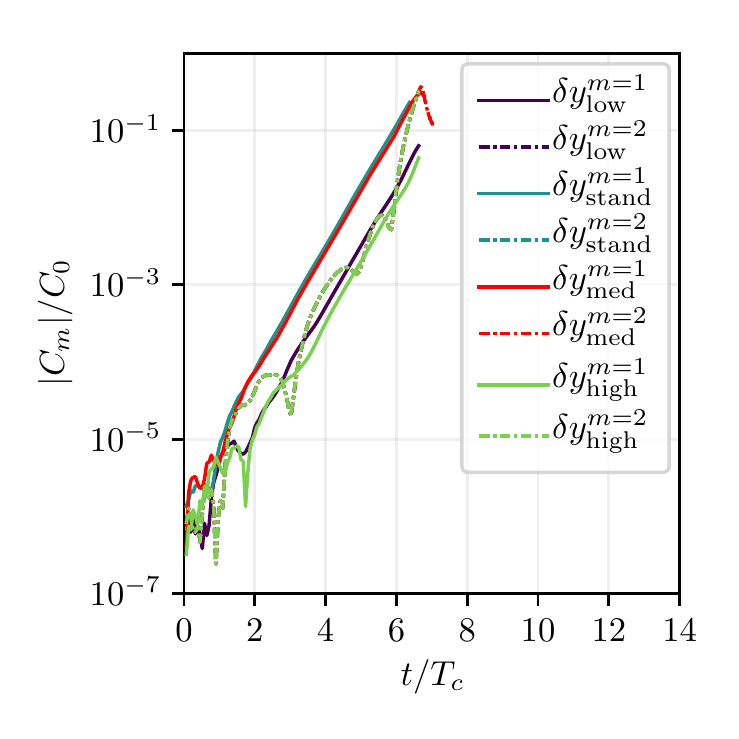} 
\caption{ 
Density mode decomposition for the
  shifted grid simulations in the case of the B model under pressure
  depletion. We show the two most dominant modes, which are always 
  the $m=1$ (solid lines) or $m=2$ (dash-dotted lines) modes. The 
  dark purple lines show the results 
  for a small grid shift $\delta y_{\rm low} = 0.0001$, the 
  blue lines shows the results for our standard simulations with grid 
  shift $\delta y_{\rm stand} = 0.001$, the red lines shows the 
  results for a medium sized grid shift $\delta y_{\rm med} = 0.005$, 
  and the green lines shows the results for a large grid shift
   $\delta y_{\rm high} = 0.01$.}
\label{fig:shift_grid}
\end{figure}

In this Appendix we discuss the results of our resolution
study  for a subset of the models presented in 
Table~\ref{tab:maxmass_prop}. We also discuss the effects of shifting the
computational grid to avoid the origin of the coordinate system.

For model A in the case of pressure depletion and model B under
both $m=1$ and $m=2$ perturbations we performed runs at 1.2 and 1.5
times the resolution of the canonical resolution discussed in 
Sec.~\ref{subsec:grid}. We find that our results are qualitatively
invariant with resolution, and that they exhibit approximate 2nd-order
convergence, which is the order of accuracy of our hydrodynamic
numerical scheme. More specifically, the dominant unstable modes
are invariant with resolution, and all collapsing models collapse to
BHs with properties that are consistent with the canonical simulations
discussed in Sec.~\ref{sec:discussion}.

We demonstrate convergence using the evolution of model B under an $m
= 2$ perturbation. In the left panel of Fig.~ \ref{fig:convergence} we
show the difference of $|\Psi^{2,2}_4|$ between the medium and
canonical resolutions, between the high and medium resolutions, and
between the high and canonical resolutions.  The curves have been
scaled assuming second order convergence, and the overlap between them
indicates approximate second order convergence.

In the center and right panels of Fig.~\ref{fig:convergence}, we show
the L2 norm of the Hamiltonian and momentum constraints, respectively,
times the squared ADM mass $M^2$ for the canonical, medium, and high
resolutions. As is clear the Hamiltonian and momentum constraints are
converging to 0 with increased resolution, and the trend is consistent
with approximate second-order convergence.

In order to avoid coordinate singularities associated with
transforming the initial data from spherical polar coordinates to
Cartesian, we shift our $y$ coordinates by a small amount to avoid the
origin of the coordinate system. To test whether our results are
affected by this choice of coordinate grids, we considered a sequence
of simulations with decreasing coordinate shift $\delta y \in [0.01,
  0.005, 0.001, 0.0001]$. We label each of these shifts as $\delta
y_{\rm high} = 0.01$, $\delta y_{\rm med} = 0.005$, $\delta y_{\rm
  stand} = 0.001$, and $\delta y_{\rm low} = 0.0001$ (note that our
standard runs employ a grid shift of $\delta y_{\rm stand}$). In
Fig.~\ref{fig:shift_grid} we show the evolution of the dominant
non-axisymmetric modes for this sequence of simulations. It is clear
that the results are practically the same for all coordinate shifts.

\bibliography{ref}

\end{document}